\documentclass[aps,amsmath,amssymb,showpacs,showkeys]{revtex4-2}
\usepackage[dvips]{graphicx}
\usepackage{times}
\usepackage{braket}
\usepackage{xcolor}
\usepackage{orcidlink}
\usepackage{mathrsfs}
\usepackage{hyperref}
\hypersetup{
	colorlinks=true,
	urlcolor=blue,
	linkcolor=red,
	citecolor=blue
}
\begin{document}
\title{Traversable wormholes in $\boldsymbol{f(Q)}$ gravity: Energy conditions, 
stability and quasinormal modes}

\author{Jaydeep Goswami\orcidlink{0009-0007-9467-5664}}
\email[Email: ]{jdpgsm98@gmail.com}

\author{Rupam Jyoti Borah\orcidlink{0009-0005-5134-0421}}
\email[Email: ]{rupamjyotiborah856@gmail.com}

\author{Umananda Dev Goswami\orcidlink{0000-0003-0012-7549}}
\email[Email: ]{umananda@dibru.ac.in}

\affiliation{Department of Physics, Dibrugarh University, Dibrugarh 786004, 
Assam, India}


\begin{abstract}
We investigate static and spherically symmetric traversable wormhole solutions 
in the framework of $f(Q)$ gravity by considering a power-law model of the 
form $f(Q)=\gamma(-Q)^m$. By adopting an anisotropic matter distribution and 
imposing an equation of state relating the radial pressure and energy density, 
we obtain an analytic shape function that satisfies the geometric requirements 
for a traversable wormhole. The model parameter is constrained to $0<m<1/2$, 
corresponding to a quintessence-like regime with $-1<\omega<-1/3$. The energy 
conditions are analyzed in detail, showing that violations of the null and weak energy conditions are unavoidable but remain localized near the wormhole 
throat. The anisotropy parameter is positive throughout the spacetime, 
indicating that repulsive anisotropic stresses play a key role in sustaining 
the wormhole. The equilibrium configuration is examined using the generalized 
Tolman-Oppenheimer-Volkoff (TOV) equation for both zero and logarithmic 
redshift functions, where a consistent force balance is achieved with 
anisotropic effects providing the dominant outward support. Dynamical 
stability is studied through scalar perturbations, leading to a 
Schr\"odinger-like wave equation with a single-peak effective potential. The 
quasinormal modes are computed using the sixth-order WKB method with Pad\'e 
approximation. The resulting frequencies possess negative imaginary parts, 
indicating stable damping of perturbations. 
Time-domain simulations further confirm the stability of the solutions and 
show good agreement with the WKB results, with small deviations in the damping 
rates. Thus, these results establish that $f(Q)$ gravity admits traversable 
wormhole solutions that are geometrically consistent and dynamically stable, 
with $f(Q)$ gravity effects effectively regulating the required matter content.
\end{abstract}



\maketitle
\section{Introduction}\label{sec01}

Wormholes are hypothetical astrophysical objects that represent tunnel-like 
structures connecting two distant regions of spacetime or even two different 
universes. These fascinating geometries arise as solutions of Einstein’s field 
equations and possess a non-trivial topology that differs fundamentally from 
ordinary flat spacetime. The concept of wormholes was first introduced by 
Wheeler in the context of spacetime quantum foam \cite{a1}, while the earliest 
mathematical representation of such structures can be traced back to the 
Einstein–Rosen bridge proposed by Einstein and Rosen in 1935 \cite{a2}. 
However, these early wormhole solutions were not traversable and were 
generally considered unstable mathematical constructs.
A significant breakthrough came with the work of Morris and Thorne \cite{a3}, 
who demonstrated that a traversable wormhole geometry could exist under 
certain physical conditions. In their formulation, the spacetime metric 
describing a static and spherically symmetric traversable wormhole contains 
two key functions: the redshift function, which determines the gravitational 
redshift experienced by signals, and the shape function, which determines the 
spatial geometry of the wormhole throat. For a physically viable wormhole 
configuration, the geometry must satisfy several conditions, such as the 
absence of horizons, the flaring-out condition at the throat, and asymptotic 
flatness \cite{a3,b1,b2,b3,b4,b5,b6,b7}.

One of the central challenges in wormhole physics is the requirement of exotic 
matter. In the framework of General Relativity (GR), the maintenance of a 
traversable wormhole requires the violation of the null energy condition 
(NEC), implying the existence of matter fields with unusual properties 
\cite{a3,a4,a5,a6,a7,a8}. Such matter that violates this energy condition 
is commonly referred to as exotic matter. The presence of exotic matter is 
problematic from a physical standpoint, motivating researchers to explore 
mechanisms that could minimize or avoid such violations. Various approaches 
have been proposed in the literature, including the use of special equations 
of state (EoSs), quantum effects, and modified as well as alternative gravity 
theories \cite{b8,b9,b10}.

In recent years, modified as well as alternative theories of gravity have 
emerged as promising frameworks for studying wormhole solutions without 
requiring large amounts of exotic matter. These theories extend Einstein’s 
gravity by modifying the geometric sector or replacing the basic geometrical 
structure of GR in the gravitational action. Former ones are usually referred 
to as modified gravity theories, whereas latter ones are referred to as 
alternative gravity theories. Several modified and alternative gravity models 
have been investigated in this context in recent years, including Rastall 
gravity \cite{a9}, $f(R)$ gravity \cite{a10,a11,a12,a13,a14,a15,a16}, 
Gauss–Bonnet gravity \cite{a17}, and teleparallel gravity \cite{a18,a19}. In 
many of these theories, additional geometric contributions effectively act as 
gravitational sources, allowing the ordinary matter threading the wormhole to 
satisfy or partially satisfy the energy conditions.

Among the various alternative gravity theories, $f(Q)$ gravity has recently 
attracted considerable attention. This theory belongs to a class of 
symmetric teleparallel gravity, where gravity is attributed to the 
non-metricity scalar $Q$ instead of curvature or torsion. The formulation of 
$f(Q)$ gravity was introduced by Jim\'enez et al.~in 2018 \cite{a20}, where 
the gravitational action is generalized from the non-metricity scalar $Q$ to 
an arbitrary function $f(Q)$. One of the notable advantages of $f(Q)$ gravity 
is that it leads to second-order field equations, similar to GR, unlike other 
modified theories, such as $f(R)$ gravity which typically involve higher-order 
derivatives. The $f(Q)$ gravity framework has been widely explored in 
cosmology and astrophysics and has been successfully applied to explain the 
accelerated expansion of the universe and has shown compatibility with 
observational data from cosmic microwave background radiation, supernovae type 
Ia and baryon acoustic oscillations \cite{a21,a22,a23,a24}. Moreover, many 
studies have investigated astrophysical objects such as black holes, 
compact stars and wormholes within this theoretical framework. For example, 
Lin and Zhai studied spherically symmetric configurations in $f(Q)$ gravity 
\cite{a25}, while Wang et al.~analysed static spherically symmetric solutions 
with anisotropic fluids \cite{a26}. Wormhole solutions supported by different 
matter sources and geometric configurations have also been studied in various 
works \cite{a27,a28,a29,a30}.

Another important aspect of wormhole physics is the choice of matter content 
described by an EoS. The EoS establishes a relation 
between pressure and energy density, and plays a crucial role in determining 
the physical characteristics of the wormhole geometry. Linear EoSs of the form 
$p_r=\omega\,\rho$ have been widely used in cosmological models and wormhole 
studies because of their simplicity and physical relevance \cite{a20}. More 
exotic fluids, such as Chaplygin gas and barotropic fluids, have also been 
investigated as possible candidates to sustain wormhole structures in 
alternative gravity theories \cite{a31}.

In addition to studying the geometric and physical properties of wormholes, it 
is also important to analyze their dynamical stability and observational 
signatures. For stability of wormhole configurations, one needs 
to study the required energy conditions as well as the various force components
that arise due to a particular matter distribution in the configuration. 
Also, it is essential to see whether the configuration can sustain over 
certain perturbations developed in the spacetime of the wormholes. 
One of the most powerful tools for probing compact objects is the 
study of quasinormal modes (QNMs), which represent the characteristic 
oscillations of spacetime resulting from perturbations \cite{a32}. These 
oscillations carry important information about the geometry and gravitational 
theory governing the system. In particular, scalar perturbations around 
wormhole backgrounds lead to wave equations with effective potentials that 
depend on the spacetime geometry. The resulting QNMs can serve as unique 
fingerprints of the underlying gravitational theory and may potentially be 
detected through gravitational wave observations in future \cite{b11}.

Motivated by these developments, the present work aims to investigate the
geometry, energy conditions, stability and scalar QNMs of Morris–Thorne 
\cite{a3} type wormhole solutions in the framework of $f(Q)$ gravity for 
two redshift functions using a range of values of the EoS parameter $\omega$. 
Accordigly the paper is organized as follows. In Section~\ref{sec02}, we 
briefly review the formulation of $f(Q)$ gravity and present the corresponding 
field equations. In Section~\ref{sec03}, we outline the general conditions for 
traversable wormhole geometries and discuss the energy conditions for 
anisotropic matter distributions. In Section~\ref{sec04}, we construct the 
wormhole solutions by adopting a power-law form of $f(Q)$ and derive the 
corresponding shape function using an EoS. The geometrical properties of the 
solutions are examined through embedding analysis. In Section \ref{sec05}, we 
analyze the physical viability of the obtained solutions by examining the 
energy conditions, including the null, weak, dominant, and strong energy 
conditions in Section. In particular, we consider two different 
wormhole configurations: a tideless wormhole with a constant redshift function 
and a wormhole with a logarithmic redshift function. Further, we investigate 
the stability of the configuration using the generalized 
Tolman-Oppenheimer-Volkoff (TOV) equation for different choices of the 
redshift function. In Section~\ref{sec06}, we study scalar perturbations of 
the wormhole spacetime and compute the corresponding QNMs using the WKB method 
with Pad\'e approximation. The behavior of the QNMs provides insights into the 
stability and observational characteristics of these wormhole geometries 
within the context of $f(Q)$ gravity. Also, we perform a 
time-domain analysis of the scalar perturbations and extract the quasinormal 
frequencies, comparing them with the WKB results. Finally, Section~\ref{sec08} 
summarizes our main results and conclusions.

\section{Framework of $f(Q)$ gravity}\label{sec02}

The action of $f(Q)$ gravity \cite{a33a,a33} is expressed as
\begin{equation}
S=\int\bigg[\frac{1}{2\kappa}f(Q)+\mathcal{L}_m\bigg]\sqrt{-g}\,d^4x, \label{eq1}
\end{equation}
where $f(Q)$ is an arbitrary function of the non-metricity scalar $Q$, $g$ is 
the determinant of the metric tensor $g_{\mu\nu}$ and $\mathcal{L}_m$ is the 
matter Lagrangian density. The spacetime in this framework is constructed by 
using the symmetric teleparallelism and non-metricity condition, i.e, 
$R^\rho{}_{\sigma\mu\nu}=0$ and $\nabla_{\alpha} \,g_{\mu \nu}\neq 0$, and 
the non-metricity tensor is defined as
\begin{equation}
Q_{\alpha\mu\nu}=\nabla_{\alpha}\,g_{\mu\nu}. 
\label{eq2}
\end{equation}
The associated affine connection can be expressed as
\begin{equation}
\Gamma^\alpha{}_{\mu\nu}=\mathring{\Gamma}^\alpha{}_{\mu\nu}+L^\alpha{}_{\mu\nu}, \label{eq3}
\end{equation}
where $\mathring{\Gamma}^\alpha{}_{\mu\nu}$ is the usual Levi-Civita 
connection, i.e.
$\mathring{\Gamma}^\alpha{}_{\mu\nu}=g^{\alpha\rho}\left(\partial_\mu g_{\rho\nu}+\partial_\nu g_{\mu\rho}+\partial_\rho g_{\mu\nu}\right)/2,$ 
and $L^\alpha{}_{\mu\nu}$ is the disformation tensor, which is defined as
\begin{equation}
L^\alpha{}_{\mu\nu} = \frac{1}{2} \left( Q^\alpha{}_{\mu\nu} - Q_\mu{}^\alpha{}_\nu - Q_\nu{}^\alpha{}_\mu \right). \label{eq5}
\end{equation}
By contracting the non-metricity tensor $Q_{\alpha\mu\nu}$ with the metric 
tensor $g_{\mu\nu}$, two independent non-metricity vectors can obtained as 
\begin{equation}
Q_\alpha = g^{\mu\nu} Q_{\alpha\mu\nu} = Q_{\alpha\mu}{}^{\mu}, \quad
\tilde{Q}_\alpha = g^{\mu\nu} Q_{\mu\alpha\nu} = Q_{\mu\alpha}{}^{\mu}.\label{eq6}
\end{equation}
On the other hand, the non-metricity scalar is defined through the contraction 
of $Q_{\alpha\beta\gamma}$ as
\begin{equation}
Q = -\,g^{\mu\nu} \left(
L^{\alpha}{}_{\beta\nu} L^{\beta}{}_{\mu\alpha}
- L^{\alpha}{}_{\beta\alpha} L^{\beta}{}_{\mu\nu}
\right)=-\,P^{\alpha\mu\nu}Q_{\alpha\mu\nu}, \label{eq7}
\end{equation}
where $P^{\alpha\mu\nu}$ is the superpotential tensor, which is symmetric in 
$\mu\nu$ and is defined as
\begin{equation}
P^{\alpha}{}_{\mu\nu}
=\frac{1}{4}\bigg[ -Q^{\alpha}{}_{\mu\nu}
+ 2\,Q_{(\mu}{}^{\alpha}{}_{\nu)}
- Q^{\alpha} g_{\mu\nu}
- \tilde{Q}^{\alpha} g_{\mu\nu}
- \delta^{\alpha}{}_{(\mu} Q_{\nu)}\bigg] . \label{eq8}
\end{equation}

Varying the action \eqref{eq1} with respect to the metric tensor $g_{\mu\nu}$ 
and considering $\kappa=1$, one can obtain the field equations for the $f(Q)$ 
gravity as
\begin{equation}
\frac{2}{\sqrt{-g}}\,\nabla_\alpha\! \left( \sqrt{-g}\, f_Q\, P^{\alpha}{}_{\mu\nu} \right)
+ \frac{1}{2}\, g_{\mu\nu} f(Q)
+ f_Q \left( P_{\mu\alpha\beta} Q_{\nu}{}^{\alpha\beta}
- 2\, Q_{\alpha\beta\mu} P^{\alpha\beta}{}_{\nu} \right)
= -\, T_{\mu\nu}, \label{eq9}
\end{equation}
where $f_Q = df(Q)/dQ$ and $T_{\mu\nu}$ is the usual matter field 
energy-momentum tensor. It is to be noted that this equation is not in a 
proper tensor form as it is only valid in the case of the coincident gauge 
coordinates \cite{a20}. However, using the affine connection defined in 
Eq.~\eqref{eq3}, one can have the curvature tensors' relation corresponding
to the present affine connection $\Gamma$ and the Levi-Civita connection
$\mathring{\Gamma}$ as given by
%
\begin{equation}
R^{\rho}{}_{\sigma\mu\nu}
= \mathring{R}^{\rho}{}_{\sigma\mu\nu}
+ \mathring{\nabla}_\mu L^{\rho}{}_{\nu\sigma}
- \mathring{\nabla}_\nu L^{\rho}{}_{\mu\sigma}
+ L^{\rho}{}_{\mu\alpha} L^{\alpha}{}_{\nu\sigma}
- L^{\rho}{}_{\nu\alpha} L^{\alpha}{}_{\mu\sigma}. \label{eq11}
\end{equation}
The corresponding relations for the Ricci tensor and Ricci scalar can
be obtained as
\begin{align}
&R_{\sigma\nu}
 = \mathring{R}_{\sigma\nu}
+ \frac{1}{2}\, \mathring{\nabla}_\nu Q_\sigma
+ \mathring{\nabla}_\rho L^{\rho}{}_{\nu\sigma}
- \frac{1}{2}\, Q_\alpha L^{\alpha}{}_{\nu\sigma}
- L^{\rho}{}_{\nu\alpha} L^{\alpha}{}_{\rho\sigma}, \label{eq12}\\[5pt]
%
&R = \mathring{R}
+ \mathring{\nabla}_\alpha Q^{\alpha}
- \mathring{\nabla}_\alpha \tilde{Q}^{\alpha}
- \frac{1}{4}\, Q_\alpha Q^{\alpha}
+ \frac{1}{2}\, Q_\alpha \tilde{Q}^{\alpha}
- L_{\rho\nu\alpha} L^{\alpha\rho\nu}. \label{eq13}
\end{align}
Now, by using the symmetric teleparallelism condition: 
$R^\rho{}_{\sigma\mu\nu}=0$, one can obtain
\begin{equation}
\mathring{R}_{\mu\nu}
- \frac{1}{2}\, \mathring{R}\, g_{\mu\nu}
= 2\, \nabla_\lambda P^{\lambda}{}_{\mu\nu}
- \frac{1}{2}\, Q\, g_{\mu\nu}
+ \left(
P_{\rho\mu\nu} Q^{\rho\sigma}{}_{\sigma}
+ P_{\nu\rho\sigma} Q_{\mu}{}^{\rho\sigma}
- 2 P_{\rho\sigma\mu} Q^{\rho\sigma}{}_{\nu}
\right). \label{eq14}
\end{equation}
Thus, we can rewrite field equations in \eqref{eq9} as \cite{a33a}
\begin{equation}
f_Q\, \mathring{G}_{\mu\nu}
+ \frac{1}{2}\, g_{\mu\nu} (Q f_Q - f)
+ 2 f_{QQ}\, \mathring{\nabla}_\lambda Q \, P^{\lambda}{}_{\mu\nu}
= T_{\mu\nu}. \label{eq15}
\end{equation}
where $f_{QQ}= df_Q/dQ$ and $\mathring{G}_{\mu\nu} = \mathring{R}_{\mu\nu}-g_{\mu\nu}\mathring{R}/2$ is the Einstein tensor. 
%
%
%

Again, varying the action \eqref{eq1} with respect to the connection, one 
may obtain other field equations of $f(Q)$ gravity as
\begin{equation}
\nabla_\mu \nabla_\nu \left( \sqrt{-g}\, f_Q\, P^{\mu\nu}{}_{\alpha} \right) = 0. \label{eq18}
\end{equation}
Nevertheless, it needs to be mentioned that for a given geometry of spacetime, 
not all $f(Q)$ models satisfy these field equations, and in the 
coincident gauge choice, these field equations behave trivially in a 
model-independent way \cite{a20,a33a}.

Although the field equations of $f(Q)$ gravity assure the conservation of the 
energy-momentum tensor as the formalism of $f(Q)$ gravity is based on its
conservation, the modification of geometry of spacetime due to the theory can 
be recast as a new energy-momentum content and hence as in most modified or
alternative gravity field equations, $f(Q)$ gravity field equations 
\eqref{eq15} can be rearranged to write in the form of Einstain field 
equations as
\begin{equation}
\mathring{G}_{\mu\nu} = \frac{T_{\mu\nu}}{f_Q} +
\frac{1}{2}\, g_{\mu\nu} \left(\frac{f(Q)}{f_Q} - Q\right)
- 2\,\frac{f_{QQ}}{f_Q}\, \mathring{\nabla}_\lambda Q \, P^{\lambda}{}_{\mu\nu}
= \frac{T_{\mu\nu}}{f_Q} + T_{\mu\nu}^{\text{\,Q}} = T_{\mu\nu}^{\,\text{eff}}, \label{eq19}
\end{equation}
where $T_{\mu\nu}^{\,\text{Q}}$ is the geometric part of the energy-momentum 
tensor in the $f(Q)$ gravity framework and $T_{\mu\nu}^{\,\text{eff}}$ is the 
effective energy-momentum tensor with contributions from both matter and 
geometry parts. The main goal of this discussion is to use the gravitational 
field equations in \eqref{eq19}, which change static and spherically symmetric 
spacetime, to analyze wormhole geometries.

\section{Conditions for traversable wormholes and the energy}
\label{sec03}

We now examine a general spherically symmetric spacetime of a static wormhole 
from the Morris-Thorne class \cite{a3, a4}. This spacetime metric can be 
expressed as
\begin{equation}
ds^2=-\,e^{2\,\Phi(r)}\,dt^2+ \left[1-\frac{b(r)}{r}\right]^{-1}\!\!\!\!dr^2+
r^2\,(d\theta^2+\sin^2\theta\,d\phi^2),  \label{eq20}
\end{equation}
where $\Phi(r)$ is the redshift function dependent on the radial coordinate 
$r$, which remains finite everywhere to avoid any event horizon. The function 
$b(r)$, called the shape function, defines the geometry of the wormhole. For 
a traversable wormhole structure,  $\Phi(r)$ and $b(r)$ must fulfill the 
following conditions:

The redshift function $\Phi(r)$ must remain finite throughout spacetime to 
prevent the formation of event horizons and ensure a well-defined temporal 
coordinate everywhere.
%
The wormhole throat occurs at the minimum radius $r=r_0$, where the shape 
function satisfies $b(r_0) = r_0$ and for $r > r_0$, the function $b(r) < r$.
%
Further, at the throat $r=r_0$, the derivative of the shape function must 
satisfy $b'(r_0) < 1$, which prevents the violation of the metric signature 
and ensures that the wormhole throat geometry is physically well-defined. 
Here, prime ($'$) denotes the derivative with respect to $r$.
%
Moreover, for a traversable geometry, the inequality $1 - b(r)/r > 0$ must be 
maintained for all $r > r_0$. This condition guarantees the absence of 
horizons and the regularity of the radial component of the metric.
%
Similarly, the spacetime should be asymptotically flat, requiring that the 
shape function satisfies $\lim\limits_{r \to \infty}b(r)/r = 0$.
%
Furthermore, to maintain a wormhole geometry that flares outward at the 
throat, the embedding condition $[b(r) - r b'(r)]/b^2(r) > 0$, evaluated near 
or at $r=r_0$, must hold. This ensures the throat forms at the minimum in the 
radial coordinate and the spacetime geometry opens outward.    

These conditions collectively define the traversable wormhole structure. 
However, additional analysis, including the violation or fulfillment of energy 
conditions and stability criteria, is required to establish the overall 
physical acceptability of the configuration. In this setup, the matter content 
within the wormhole is represented by an anisotropic energy-momentum tensor 
as follows:
\begin{equation}
T^{\nu}_{\mu}
= (\rho + p_t) u_\mu u^\nu
- p_t \delta^\nu_\mu
+ (p_r - p_t) v_\mu v^\nu, \label{eq21}
\end{equation}
where $u_\mu$ is the four-velocity, $v_\mu$ is a unit spacelike vector in the 
radial direction, $\rho$ is the energy density, and $p_r$ and $p_t$ are the 
radial and tangential pressures respectively, which are functions of the 
radial coordinate $r$.
%
%
Energy conditions are a set of mathematical constraints that characterize the 
physical behavior of matter and energy in spacetime, and are closely related 
to the Raychaudhuri equation~\cite{a34}. For the present setup, the standard 
energy conditions are defined as follows:
\begin{itemize}
\item Null Energy Condition (NEC): For any null vector $k^\mu$, the 
effective energy-momentum tensor $T_{\mu\nu}^{\,\text{eff}}$ satisfies
$T_{\mu\nu}^{\text{\,eff}}\,k^\mu k^\nu \geq 0$ with $k^\mu k_\mu = 0.$
For an anisotropic fluid, this reduces to
$\rho^{\text{\,eff}} + p_r^{\text{\,eff}} \geq 0$ and 
$\rho^{\text{\,eff}} + p_t^{\text{\,eff}} \geq 0.$
Violation of the NEC typically indicates the presence of exotic matter 
required to sustain a wormhole throat.

\item Weak Energy Condition (WEC): For any timelike vector $u^\mu$, 
$T_{\mu\nu}^{\,\text{eff}}$ has to satisfy
$T_{\mu\nu}^{\text{\,eff}}\,u^\mu u^\nu \geq 0$ with $u^\mu u_\mu = -\,1.$
For an anisotropic fluid, this condition implies
$\rho^{\text{\,eff}} \geq 0,$
$\rho^{\text{\,eff}} + p_r^{\text{\,eff}} \geq 0$ and
$\rho^{\text{\,eff}} + p_t^{\text{\,eff}} \geq 0.$
The WEC ensures that all observers measure a non-negative energy density.

\item Strong Energy Condition (SEC): For any timelike vector $u^\mu$,
$\left(T_{\mu\nu}^{\text{\,eff}} - 
\frac{1}{2}\, g_{\mu\nu} T^{\text{\,eff}}\right) u^\mu u^\nu \geq 0,$ 
where $T^{\text{\,eff}} = g^{\mu\nu} T_{\mu\nu}^{\text{\,eff}}$. For an 
anisotropic fluid, this gives
$\rho^{\text{\,eff}} + p_r^{\text{\,eff}} + 2\,p_t^{\text{\,eff}} \geq 0.$
The SEC is associated with the focusing of timelike geodesics.

\item Dominant Energy Condition (DEC): For any timelike vector $u^\mu$,
$T_{\mu\nu}^{\text{\,eff}}\,u^\mu u^\nu \geq 0$ with
$T_{\mu\nu}^{\text{\,eff}}\,u^\mu < 0.$ For an anisotropic fluid, this leads 
to $\rho^{\text{\,eff}} - |p_r^{\text{\,eff}}| \geq 0$ and
$\rho^{\text{\,eff}} - |p_t^{\text{\,eff}}|\geq 0.$
\end{itemize}
In GR, physically reasonable matter fields are expected to satisfy these 
energy conditions. The study of wormholes gained significant attention 
following a landmark paper by Morris et al.~\cite{a3}, which introduced a 
spherically symmetric metric for a traversable wormhole. However, this 
wormhole solution necessitates the presence of exotic matter, leading to a 
violation of the NEC. In the framework of modified theories of gravity, 
higher-order curvature terms in the field equations or new spacetime 
structures in alternative gravity frameworks can act as effective 
sources of gravitational energy, modifying the energy-momentum structure and 
potentially relaxing or even bypassing these violations \cite{a35}.

In passing, it needs to be mentioned that the anisotropy parameter refers to 
a quantity that measures the degree of anisotropy or directional dependence 
of a physical system or spacetime geometry. For the present case, it can be 
defined as
\begin{equation}
\Delta = p_t^{\text{\,eff}} - p_r^{\text{\,eff}}.
\label{eq22}
\end{equation}
The geometry is attractive for $\Delta < 0$ and repulsive for $\Delta > 0$, 
while $\Delta = 0$ corresponds to an isotropic fluid configuration.

\section{Wormhole geometry in $f(Q)$ gravity}\label{sec04}
From Eq.~\eqref{eq7}, the non-metricity scalar $Q$ for the wormhole metric 
\eqref{eq20} is calculated as
\begin{equation}
Q=-\frac{2}{r} \bigg[1-\frac{b(r)}{r}\bigg] \bigg[2\Phi'(r)+\frac{1}{r}\bigg]. \label{eq23}
\end{equation}
Since the calculation of the non-metricity scalar $Q$ is lengthy, we present 
the details, along with the corresponding non-metricity tensor and 
superpotential tensor, in the Appendix. From the definition of 
$T_{\mu\nu}^{\text{\,Q}}$ in Eq.~\eqref{eq19}, we get
\begin{align}
\rho^{\text{Q}} =\,&
-\frac{f(Q)}{2 f_Q}
- \frac{(r - b(r))\left(1 + 2 r \Phi'(r)\right)}{r^3} \notag\\
& + \frac{4 f_{QQ} (r - b(r))}{f_Q r^6}
\bigg[
b(r)\left(-3 - 4 r \Phi'(r) + 2 r^2 \Phi''(r)\right)
\nonumber \\
& + r \Big( 2 + b'(r)\left(1 + 2 r \Phi'(r)\right)
+ 2 r \left(\Phi'(r) - r \Phi''(r)\right)\!\!\Big)
\bigg],
\label{eq24}\\[8pt]
%
p_r^{\text{Q}} =\,& \frac{f(Q)}{2 f_Q} + \frac{1}{r^2}
- \frac{b(r)}{r^3}
+ \frac{2\bigl(r - b(r)\bigr)\Phi'(r)}{r^2}, \label{eq25}\\[8pt]
%
p_t^{\text{Q}} =\;&
\frac{f(Q)}{2 f_Q}
+ \frac{(r - b(r))\left(1 + 2r\,\Phi'(r)\right)}{r^3}
\nonumber \\
&+ \frac{2 f_{QQ}\,(r - b(r))\left(1 + r\,\Phi'(r)\right)}{f_Q\,r^6}\bigg[
b(r)\left(3 + 4r\,\Phi'(r) - 2r^2\,\Phi''(r)\right)
\nonumber \\
&
+ r\Big(\!-2 - b'(r) - 2r\big(1 + b'(r)\big)\Phi'(r)
+ 2r^2\,\Phi''(r)\!\Big)\bigg].
\label{eq26}
\end{align}
Similarly, for the metric \eqref{eq20} the non-zero components of the Einstein 
tensor $\mathring{G}^{\mu}_{\nu}$ can be obtained as
\begin{align}
\mathring{G}^{t}_{t} & = -\,\frac{b'(r)}{r^2}, \label{eq27}\\[5pt]
\mathring{G}^{r}_{r} & = \frac{2\,r\,\Phi'(r)\left(r - b(r)\right)-b(r)}{r^3}, \label{eq28}\\[5pt]
\mathring{G}^{\theta}_{\theta} & = \mathring{G}^{\phi}_{\phi} =\frac{\left(1 + r\,\Phi'(r)\right)\left(b(r) - r\,b'(r) + 2r\left(r - b(r)\right)\Phi'(r)\right)}{2r^3}
+ \frac{\left(r - b(r)\right)\Phi''(r)}{r}. \label{eq29}
\end{align}

From Eqs.~\eqref{eq24} - \eqref{eq26} it is clear that to perform the 
investigation on the shape-function and energy conditions of the wormhole in 
the background of $f(Q)$ gravity scenario, we need a specific form of the
function $f(Q)$ and hence for this work we consider the power-law form of the
model or function as given by~\cite{a36}
\begin{equation}
f(Q)=\gamma\,(-Q)^m, \label{eq30}
\end{equation}
where $\gamma$ and $m$ are constants. It should be noted that as $b(r)<r$, 
the non-metricity scalar $Q(r)$ always remains negative and hence the minus 
sign appears in the function \eqref{eq30} to avoid the negativity of the 
model. Using the model \eqref{eq30} and the anisotropic energy-momentum tensor 
$T^{\mu}_{\nu}$ in \eqref{eq21}, we can calculate the energy density 
$\rho$, the radial pressure $p_r$ and the tangential pressure $p_t$, 
respectively, for the $T_{\mu\nu}/f_Q$ part of field equations in \eqref{eq19} 
as
\begin{align}
\rho =\,&
\frac{2^{\,m-1}\,\gamma}{(b(r)-r)\left(1 + 2r\,\Phi'(r)\right)^2}
\left[\frac{(r - b(r))\left(1 + 2r\,\Phi'(r)\right)}{r^3}\right]^m
\nonumber \\
&\times \bigg\{
m(2 m - 1)\, r\, b'(r)\left(1 + 2r\,\Phi'(r)\right) + (m - 1)\Big[
r(1 + 4 m)- b(r)(1 + 6 m)
\nonumber \\
&
+ 4r\Big(
\Phi'(r)\big(r(1 + m) - b(r)(1 + 2 m) + r(r - b(r))\Phi'(r)\big)+ m(r - b(r))\, r\, \Phi''(r)
\Big)
\Big]
\bigg\},
\label{eq31}\\[8pt]
%
p_r =\,&
-\frac{2^{m-1}\,\gamma}{\left(b(r) - r\right)\left(1 + 2r\,\Phi'(r)\right)}
\left[\frac{(r - b(r))\left(1 + 2r\,\Phi'(r)\right)}{r^3}\right]^{m}\!\! \left[
b(r) + (m - 1)r + 2\bigl(b(r) - r\bigr)r\,\Phi'(r)
\right],
\label{eq32}
\end{align}
\begin{align}
p_t =&
-\frac{m\,\gamma}{r^3}\, 2^{m-2}
\left[
\frac{(r - b(r))\left(1 + 2r\,\Phi'(r)\right)}{r^3}
\right]^{m-1}
\nonumber \\
&\times \bigg\lbrace
2(b(r) - r)\left(1 + 2r\,\Phi'(r)\right)
+ \left(1 + r\,\Phi'(r)\right)
\left(b(r) - r\,b'(r) + 2r(r - b(r))\,\Phi'(r)\right)
\nonumber \\
&
- \frac{2(b(r) - r)\left(1 + 2r\,\Phi'(r)\right)}{m}
+ 2r^2 (r - b(r))\,\Phi''(r) - \frac{2(m-1)\left(1 + r\,\Phi'(r)\right)}{1 + 2r\,\Phi'(r)}
\nonumber \\
&
\Big(\!\!
-3b(r) + 2r
+ r\,b'(r)\left(1 + 2r\,\Phi'(r)\right)
+ 2r\left((r - 2b(r))\,\Phi'(r)
+ (b(r) - r)r\,\Phi''(r)\right)\!\Big)
\bigg\rbrace.
\label{eq33}
\end{align}
%

To solve Eqs.~\eqref{eq31} and \eqref{eq32} for the shape function $b(r)$ of 
the wormhole, we consider an equation of state (EoS), i.e., a relation between 
the radial pressure and the energy density of the matter distribution as 
given by  
\begin{equation}
p_r = \omega\,\rho, \label{eq34}
\end{equation} 
where $\omega$ is the EoS parameter that characterizes the nature of the 
matter content. Substituting equations \eqref{eq31} and \eqref{eq32} in 
this EoS \eqref{eq34} for the tideless condition, i.e., for 
$\Phi(r)=\Phi'(r)=0$, we get
\begin{equation}
b'(r) = \frac{b(r)\,\big[(m-1)(6 m+1)\,\omega -1\big]-\big[(m-1)\,r\,\big(1+\omega(1+4 m)\big)\big]}{m\,\omega\,(2 m - 1)\,r}. \label{eq35}
\end{equation} 
This equation relates the geometry to the matter content and allows us to 
solve for the shape function. Integrating this Eq.~\eqref{eq35} yields the  
general form of the shape function, which is given by
\begin{equation}
b(r)= \frac{B}{A+1}\,r + C\,r^{-A}; \;\; \text{for}\;\; (A \neq -1), 
\label{eq36}
\end{equation}
where $A$ and $B$ are constants and have the forms:
\begin{equation}
A = \frac{1+\left(1-m\right)\,\left(6m+1\right)\,\omega}{m\,\omega\left(2\,m-1\right)}
\quad \text{and} \quad
B =\frac{\left(1-m\right)\,\left(1+\omega(1+ 4m)\right)}{m\,\omega\,\left(2m-1\right)}, \label{eq37}
\end{equation}
and $C$ is an integration constant. Using the throat condition $b(r_0)=r_0$, 
we get $C= r_0^{A+1}\,(1- B/(A+1))$, and hence the shape function~\eqref{eq36} 
takes the form:
\begin{equation}
b(r)= \frac{B}{A+1}\,r + \left(1-\frac{B}{A+1}\right)\,r_0\,\left(\frac{r_0}{r}\right)^{A}\!\!\!. \label{eq38}
\end{equation}
The asymptotic flatness condition $\lim\limits_{r \to \infty}b(r)/r = 0$ 
implies that $A > -1$ and the linear term vanishes, i.e., $B=0$ as 
$r \to \infty$. This yields the expression of the EoS parameter as
\begin{equation}
\omega = -\, \frac{1}{4m+1}\,, \label{eq39}
\end{equation}
if $m\neq 1$. Using these conditions, the shape function becomes
\begin{equation}
b(r)=r_0^{A+1}\,r^{-A}, \quad \text{with}\; A=-\,\frac{6m-1}{2\,m-1}. 
\label{eq40}
\end{equation}
Applying the flare-out condition $b'(r)<1$, we get $0<m<0.5$. Using this range 
of the parameter $m$, Eq.~\eqref{eq39} gives $-1<\omega<-1/3$. 
Therefore, the wormhole is supported by a geometry that mimics exotic
but non-phantom matter in the quintessence regime. The negative radial pressure 
(tension) is essential to sustain the wormhole throat against gravitational 
collapse, while remaining within the quintessence regime.
%

\begin{figure}[!h]
    \includegraphics[scale=0.65]{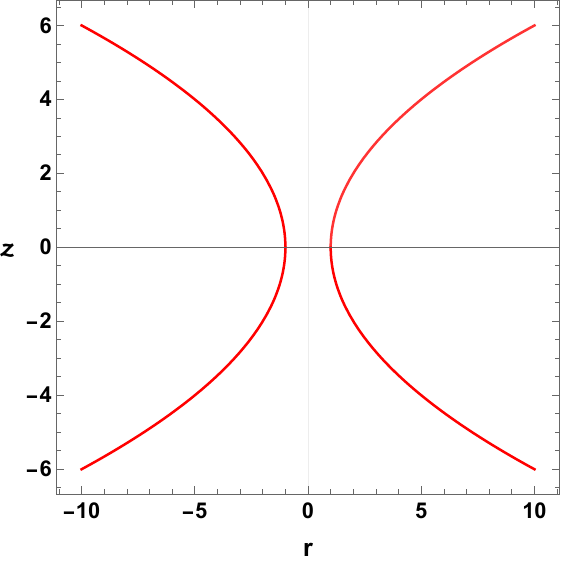}\hspace{1.0cm}
    \includegraphics[scale=0.7]{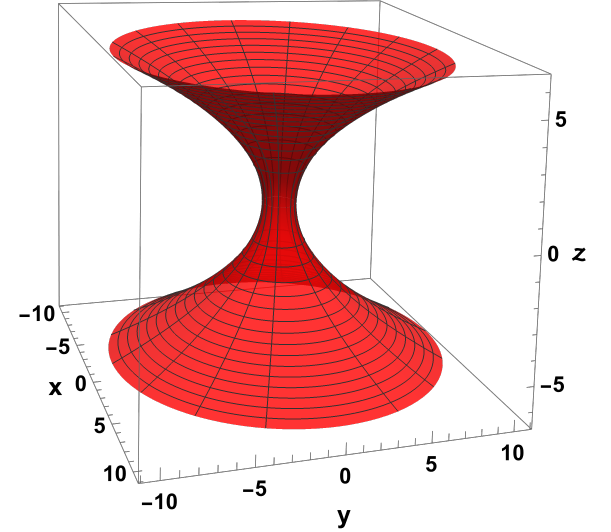}
    \vspace{-0.2cm}
    \caption{2-D (left) and 3-D (right) embedded plots of the wormhole defined 
by the shape function \eqref{eq40} with $A=0$ and the throat radius $r_0 = 1$.}
    \label{fig1}
\end{figure}

In order to visualise the embedded diagrams of the wormholes represented by 
the shape function~\eqref{eq40}, we consider an equatorial slice 
$\theta = \pi/2$ at a fixed time or at $t =$ constant. This consideration 
reduces the wormhole metric \eqref{eq20} into the form:
\begin{equation}
ds^2=\left[1-\frac{b(r)}{r}\right]^{-1}\!\!\!\!dr^2+r^2\,d\phi^2.  \label{eq41}
\end{equation}
For the considered setting as stated above, in cylindrical coordinates, one 
can write the spacetime metric as
\begin{equation}
ds^2= dz^2+dr^2+r^2\,d\phi^2. \label{eq42}
\end{equation}
Since in three dimensional Euclidean space the embedded surface is defined by 
$z = z(r)$, Eq.~\eqref{eq42} can be rewritten as
\begin{equation}
ds^2=\left[1 + \left(\frac{dz}{dr}\right)^2\right]\,dr^2 + r^2\,d\phi^2. 
\label{eq43}
\end{equation}
Comparing this redefined metric \eqref{eq43} with the reduced metric 
\eqref{eq41}, we get
\begin{equation}
\frac{dz}{dr}= \pm\, \sqrt{\frac{b(r)}{r - b(r)}}. \label{eq44}
\end{equation}
This relation can be used to obtain the wormhole's embedded surface. Moreover, 
from the flare-out condition, it can be seen that at the throat of the 
wormhole defined by the shape function~\eqref{eq40}, the inverse of the 
embedding function, i.e., $r(z)$ satisfies the condition $d^2 r/dz^2 > 0$. 
Specifically, differentiation of the inverse of Eq.~\eqref{eq44} with respect 
to $z$ leads to the condition:
\begin{equation}
\frac{d^2 r}{dz^2}\Bigg|_{r\,=\,r_0}\!\!\!\!\! = \frac{b(r_0) - r_0 b'(r_0)}{2 b(r_0)^2} > 0 \label{eq45}
\end{equation}
Furthermore, one can see from Eq.~\eqref{eq44} that at the wormhole throat 
$dz/dr \to \infty$, and as $r \to \infty$ the wormhole space becomes 
asymptotically flat. Again, substituting the shape function~\eqref{eq40} in 
Eq.~\eqref{eq44}, we get 
\begin{equation}
z = \pm \int_{r_0}^{r}\left[\left(\frac{r_0}{r}\right)^{A+1} - 1\right]^{-\,1/2}\!\!\!\! dr.
\label{eq46}
\end{equation}
For $A=0$, corresponding to $m=1/6$ or $\omega=-\,3/5$, the shape 
function in Eq.~\eqref{eq40} reduces to a constant, $b(r) = r_0$, which 
represents a simple zero-tidal-force wormhole. Consequently, Eq.~\eqref{eq46} 
yields the embedding function $z(r) = \pm\, 2\sqrt{r_0\left(r - r_0\right)}$, 
indicating a parabolic geometry near the throat. Fig.~\ref{fig1} illustrates 
the corresponding embedded diagrams of the wormholes. The embedding diagram 
exhibits two symmetric branches corresponding to the two asymptotically flat 
regions connected by the wormhole throat $r_0$ and the surface becomes 
vertical at $r=r_0$, indicating the flare-out condition is satisfied. The 
smooth and symmetric parabolic embedding reflects the absence of tidal 
forces, consistent with the tideless condition ($\Phi(r) = 0$).

\begin{figure}[!h]
    \includegraphics[scale=0.75]{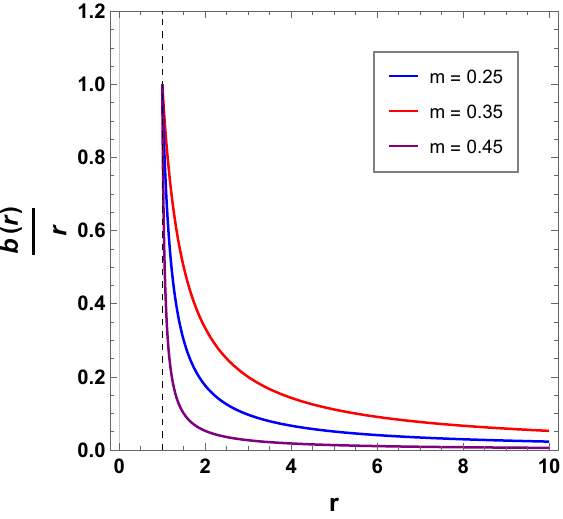}
    \hspace{1.0cm}
    \includegraphics[scale=0.7]{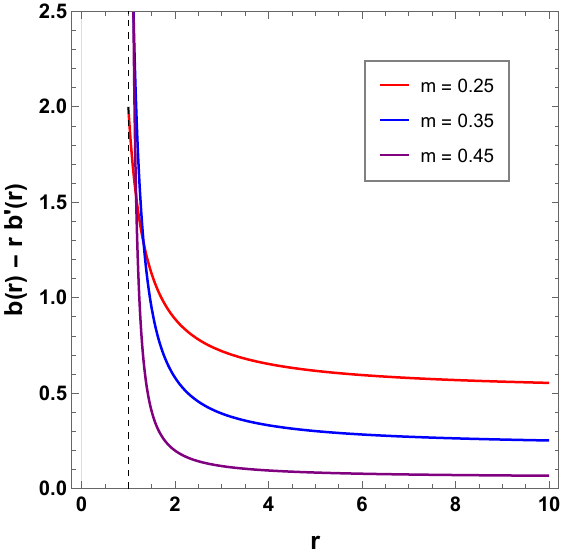}
    \vspace{-0.2cm}
    \caption{Plots of functions $b(r)/r$ versus $r$ (left panel) and 
$b(r)-r b'(r)$ versus $r$ (right panel) for the wormhole shape function 
\eqref{eq49} with different values of the parameter $m$ and throat radius 
$r_0 = 1$. The vertical dotted line in each plot represents the position of 
the throat of the wormhole.}
    \label{fig2}
\end{figure}

\begin{figure}[!h]
    \includegraphics[scale=0.67]{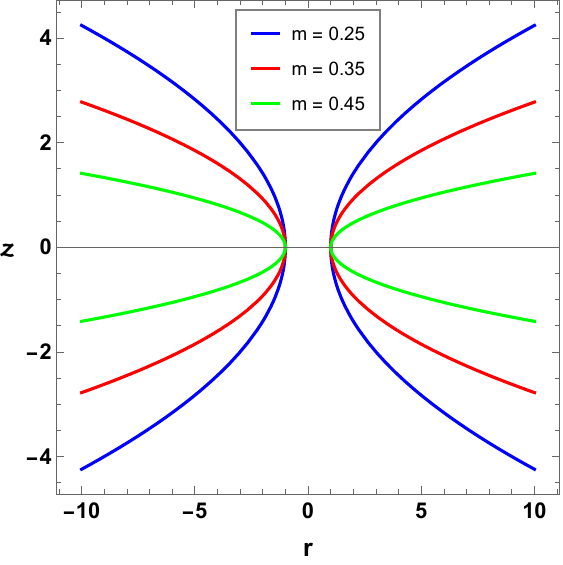}
    \hspace{1.0cm}
    \includegraphics[scale=0.55]{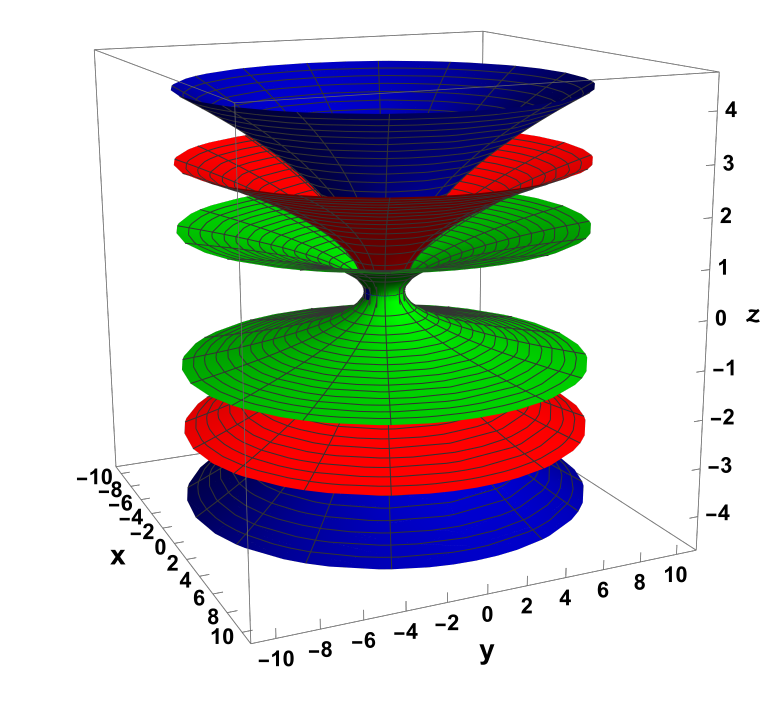}
    \vspace{-0.2cm}
    \caption{2-D (left panel) and 3-D (right panel) embedded plots of the 
wormhole defined by the shape function~\eqref{eq49} with different values of 
the parameter $m$ and throat radius $r_0 = 1$. The blue, red and green plots on 
the right plot correspond to $m=0.25$, $m=0.35$ and $m=0.45$, respectively.}
    \label{fig3}
\end{figure}

However, for general $r > r_0$, the integral in Eq.~\eqref{eq46} leads to 
an expression in terms of hyper-geometric functions, making it analytically 
intractable. Therefore, to obtain a tractable form near the throat, we perform 
a Taylor series expansion of the shape function about $r = r_0$ by setting 
$r = r_0 + \epsilon$, where $\epsilon \ll r_0$. The shape function then takes 
the form:
\begin{equation}
b(r) = b(r_0) + b'(r_0)(r - r_0) + \mathcal{O}(\epsilon^2). \label{eq47}
\end{equation}
Using the near-throat approximation obtained above, equation \eqref{eq44} 
reduces to
\begin{equation}
\frac{dz}{dr} \approx \sqrt{\frac{r_0}{(A+1)(r-r_0)}}. \label{eq48}
\end{equation}
Comparing this with Eq.~\eqref{eq44}, we obtain the following form of the 
shape function:
\begin{equation}
b(r) = \frac{r\,r_0}{A(r - r_0) + r}. \label{eq49}
\end{equation}
To examine the physical viability of the shape function~\eqref{eq49}, we 
analyse the necessary wormhole conditions, namely the throat condition, 
asymptotic flatness and the flare-out condition. For this purpose, the 
functions $b(r)/r$ and $b(r)-r b'(r)$ are plotted with respect to $r$, along 
with the corresponding embedding diagrams obtained from Eq.~\eqref{eq48}, as 
shown in Figs.~\ref{fig2} and \ref{fig3}, respectively.

From Fig.~\ref{fig2} (left panel), it is seen that the ratio $b(r)/r$ is a 
monotonically decreases monotonically as $r$ increases and tends to zero as $r 
\to \infty$, confirming the asymptotic flatness of the spacetime for all 
considered values of $m$. Moreover, larger values of $m$ lead to a faster decay 
of $b(r)/r$, which indicates a more rapid approach to asymptotic flatness. In 
the right panel, the quantity $b(r) - r b'(r)$ remains strictly positive for $r 
\ge r_0$, thereby satisfying the flare-out condition required for a traversable 
wormhole. The function attains its maximum near the throat $r=r_0$ and 
decreases monotonically with increasing $r$, indicating that the flaring of 
the geometry is strongest at the throat and gradually gets weaker for $r > r_0$. 
This behavior is consistent with the transition to an asymptotically flat 
spacetime. Furthermore, increasing $m$ results in a more rapid decay of the 
function, which implies that the wormhole geometry becomes increasingly 
localized around the throat.
The embedding diagrams in Fig.~\ref{fig3} further support these results. In 
the 2-D embedding (left panel), the geometry is symmetric about $z=0$, that
represents two identical asymptotically flat regions connected through the 
throat. The surface smoothly flares outward from the throat, thereby confirming 
the traversable nature of the wormhole. As $m$ increases, the extent of the 
embedding surface along the $z$-direction decreases which indicates a more compact geometry. This feature is also seen in the 3-D embedding plot (right panel), where larger values of $m$ produce surfaces that are more confined near the throat, while smaller values of $m$ correspond to more extended geometries. This demonstrates that the parameter $m$ controls the spatial curvature and effective size of the wormhole.

Therefore, the shape function~\eqref{eq49} satisfies all the essential 
conditions for a physically viable traversable wormhole, as confirmed by the 
asymptotic behavior, flare-out condition, and embedding analysis, and will be 
used for subsequent investigations.

\begin{figure}[!h]
    \includegraphics[scale=0.85]{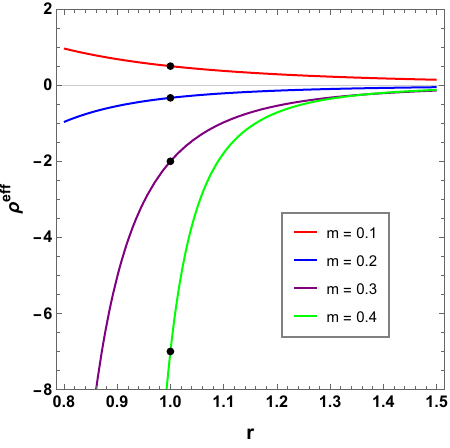}
    \vspace{-0.2cm}
    \caption{Variation of effective energy density $\rho^{\,\text{eff}}$ with
respect to radial coordinate $r$ of the wormhole defined by the shape 
function~\eqref{eq49} with different values of the parameter $m$ and throat 
radius $r_0 = 1$.}
    \label{fig4}
\end{figure}

Using the shape function~\eqref{eq49} and Eqs.~\eqref{eq31} - \eqref{eq33}, 
we can obtain the effective energy density, radial pressure and tangential 
pressure from the field equations in \eqref{eq19} as
\begin{align}
\rho^{\,\text{eff}} =\, &
\frac{\left(12m^2 - 8m + 1\right) r_0^2}
{r^2 \left(4mr - 6mr_0 + r_0\right)^2}, \label{eq50}\\[8pt]
p_r^{\,\text{eff}} =\, &
\frac{8 m r (r - r_0)\, \Phi'(r) + (2 m - 1) r_0}{r^2 \left(4 m r - 6 m r_0 + r_0 \right)}, \label{eq51}\\[8pt]
p_t^{\,\text{eff}} = \,& \frac{4 m r (r - r_0)\, \Phi''(r)\, \left(4 m r - 6m r_0 + r_0 \right) + 2 m \left(r \Phi'(r) + 1 \right)}
{r \left(4m r - 6m r_0 + r_0 \right)^2}\notag \\
& \times \Big( 2 (r - r_0)\, \Phi'(r)\, \left(4 m r - 6m r_0 + r_0 \right) - 2 m r_0 + r_0 \Big). 
\label{eq52}
\end{align}
From Eq.~\eqref{eq50}, it is clear that the effective energy density is 
independent of the redshift function $\Phi(r)$. Therefore, the same expression 
applies to both wormhole configurations, with and without a redshift function. 
We plot Eq.~\eqref{eq50} as a function of the radial coordinate $r$ in 
Fig.~\ref{fig4}. In the plot, the black dots denote the values of the 
corresponding quantities at the throat radius $r=r_0$.
In Fig.~\ref{fig4}, it is observed that the effective energy density 
$\rho^{\,\text{eff}}$ exhibits a strong dependence on the model parameter $m$. 
In particular, $\rho^{\,\text{eff}}$ becomes negative at the wormhole throat 
for $1/6 < m < 1/2$, indicating a violation of WEC in this range, and  
hence, indicates the presence of exotic matter in this parameter range. For 
smaller values of $m$, the energy density remains positive. Furthermore, 
$\rho^{\,\text{eff}}$ approaches zero as $r$ increases, indicating that the 
violations are localized around the wormhole throat, which is a desirable 
feature for physically viable wormhole models.

\section{Energy Conditions and Stability Analysis in Anisotropic Fluid 
Configuration}\label{sec05}

In this section, we examine the energy conditions for the anisotropic matter 
distribution supporting the wormhole geometry. By treating the radial and 
tangential pressures separately, we analyze the null, weak, strong, and 
dominant energy conditions in terms of the effective energy density and 
pressure components. Particular attention is given to the role of the model 
parameter $m$, which governs the behavior of the energy density and pressures. 
Using analytical and graphical analysis, we investigate how different values 
of $m$ affect the satisfaction or violation of the energy conditions, thereby 
assessing the physical plausibility of the proposed shape 
function~\eqref{eq49}. Furthermore, we analyze the anisotropy parameter 
defined in Eq.~\eqref{eq22} and study the stability of the system via the 
Tolman-Oppenheimer-Volkoff (TOV) equation~\cite{a37}. To examine the 
mechanical equilibrium of the wormhole supported by an anisotropic fluid 
distribution, we employ the generalized TOV equation given by
\begin{equation}
\frac{d p_r^{\,\text{eff}}}{dr} 
+ \frac{2}{r}\left(p_r^{\,\text{eff}} - p_t^{\,\text{eff}}\right) 
+ \frac{d\Phi(r)}{dr}\left(\rho^{\,\text{eff}} + p_r^{\,\text{eff}}\right) = 0. \label{eq53}
\end{equation}
This equation can be decomposed into three distinct force components:
hydrostatic force: $F_H = -\,d p_r^{\,\text{eff}}/dr$,
anisotropic force: $F_A = 2\left(p_t^{\,\text{eff}} - p_r^{\,\text{eff}}\right)/r$,
gravitational force: $F_G = -\,(d\Phi/dr)\left[\rho^{\,\text{eff}} + p_r^{\,\text{eff}}\right]$.
%
The wormhole remains in equilibrium when these forces balance each other, i.e.,
\begin{equation}
F_H + F_A + F_G = 0. \label{eq54}
\end{equation}

For our analysis, we consider two choices for the redshift function: (i) zero 
redshift function, i.e., $\Phi(r)=0$, and (ii) a logarithmic 
redshift function, $\Phi(r)=\log\left(1+r_0/r\right)$. For the case of 
$\Phi(r) = 0$ or $e^{2 \Phi(r)} = 1$, the wormhole is said to be tideless
\cite{a38}. This simplifies the calculations associated with the field 
equations and provides interesting wormhole solutions. The later case, i.e., 
the logarithmic redshift function, which remains finite throughout the 
spacetime, including at the throat $r=r_0$, ensures that the metric 
coefficient $g_{tt}$ remains non-vanishing and thereby avoids the formation 
of event horizons. Also, this choice has been widely adopted in previous 
studies of wormhole geometries (e.g., see Refs.~\cite{a14,a39,a40,a41,a42}). 

\subsection{For $\boldsymbol{\Phi(r)=0}$ case}\label{subsec0501}
\begin{figure}[!h]
    \centerline{
    \includegraphics[scale=0.85]{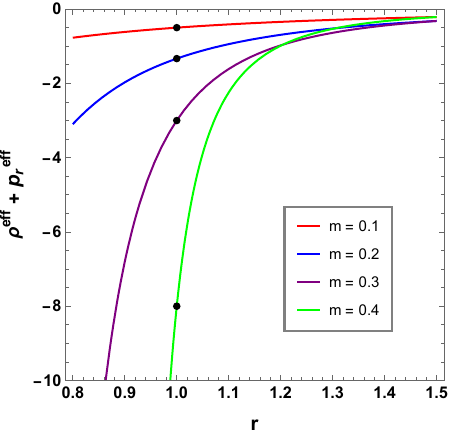}
    \hspace{1cm}
    \includegraphics[scale=0.825]{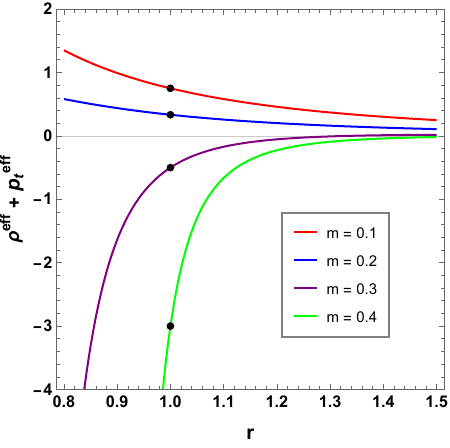}}
    \vspace{-0.2cm}
    \caption{Plots of $\rho^{\,\text{eff}} + p_r^{\,\text{eff}}$ (left panel) 
and $\rho^{\,\text{eff}} + p_t^{\,\text{eff}}$ (right panel) with respect to 
radial coordinate $r$ for different values of the parameter $m$, throat 
radius $r_0 = 1$ and $\Phi(r)=0$.}
    \label{fig5}
\end{figure}

\begin{figure}[!h]
    \centerline{
    \includegraphics[scale=0.85]{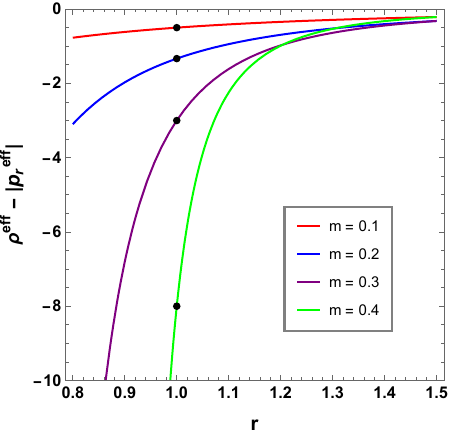}
    \hspace{1cm}
    \includegraphics[scale=0.85]{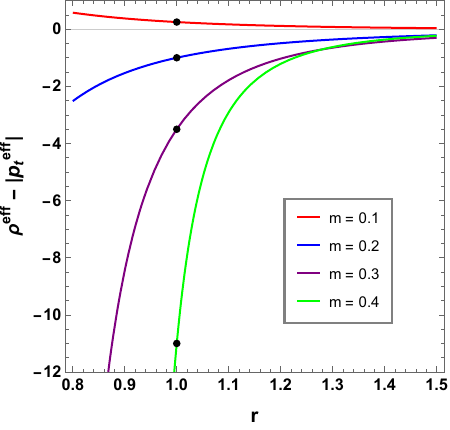}}
    \vspace{-0.2cm}
    \caption{Plots of $\rho^{\,\text{eff}} - \left|p_r^{\,\text{eff}}\right|$ 
(left panel) and $\rho^{\,\text{eff}} - \left|p_t^{\,\text{eff}}\right|$ 
(right panel) with respect to radial coordinate $r$ for different values of 
the parameter $m$, throat radius $r_0 = 1$ and $\Phi(r)=0$.}
    \label{fig6}
\end{figure}

%
%
For this tideless condition, from Eqs.~\eqref{eq51} and \eqref{eq52}, we can 
obtain the specific forms of effective radial pressure and tangential pressure 
as
\begin{align}
p_r^{\text{eff}} & =
\frac{(2m - 1)\, r_0}{r^2 \left(4mr - 6mr_0 + r_0\right)}, \label{eq56}\\[8pt]
%
p_t^{\text{eff}} & = 
\frac{2m(1 - 2m)\, r_0}
{r \left(4mr - 6mr_0 + r_0\right)^2}. \label{eq57}
\end{align}
Using these Eqs.~\eqref{eq56} and~\eqref{eq57}, the specific expressions for 
the four energy conditions can be found as
\begin{align}
\rho^{\,\text{eff}} + p_r^{\,\text{eff}} & = \frac{4m(2m - 1)\, r_0}
{r \left(4mr - 6mr_0 + r_0\right)^2}, \label{eq58}\\[8pt]
%
\rho^{\,\text{eff}} + p_t^{\text{\,eff}} & =
-\frac{(2m - 1)\, r_0 \left[2m(r - 3r_0) + r_0\right]}
{r^2 \left(4mr - 6mr_0 + r_0\right)^2}, \label{eq59}\\[8pt]
%
\rho^{\,\text{eff}} + p_r^{\,\text{eff}} & + 2  p_t^{\,\text{eff}} =0, \label{eq60}\\[8pt]
%
\rho^{\,\text{eff}} - \left|p_r^{\,\text{eff}}\right|& =
\frac{\left(12m^2 - 8m + 1\right) r_0^2}
{r^2 \left(4mr - 6mr_0 + r_0\right)^2}
-
\left|
\frac{(2m - 1)\, r_0}
{r^2 \left(4mr - 6mr_0 + r_0\right)}
\right|, \label{eq61}\\[8pt]
%
\rho^{\,\text{eff}} - \left|p_t^{\,\text{eff}}\right|
& =
\frac{\left(12m^2 - 8m + 1\right) r_0^2}
{r^2 \left(4mr - 6mr_0 + r_0\right)^2}
-
\left|
\frac{2m(1 - 2m)\, r_0}
{r \left(4mr - 6mr_0 + r_0\right)^2}
\right|. \label{eq62}
\end{align}
From Eq.~\eqref{eq60}, it is observed that effective SEC is satisfied 
within the model parameter $m$ range $0 < m < 1/2$.
To examine the validity of the energy conditions more explicitly, we plot 
Eqs.~\eqref{eq58}, \eqref{eq59}, \eqref{eq61}, and~\eqref{eq62} as functions 
of the radial coordinate $r$, as shown in Figs.~\ref{fig5}, and \ref{fig6}.

Fig.~\ref{fig5} shows that the radial NEC remains negative throughout the 
spacetime for all values of $m$, indicating a persistent violation in the 
radial direction, while the tangential NEC is satisfied for smaller values of 
$m$, but becomes negative for $1/4 < m < 1/2$ at the throat, showing that the 
violation in the tangential direction occurs only within a restricted 
parameter range. Also, WEC is violated for all considered values of $m$. This 
is due to the fact that $\rho^{\,\text{eff}} + p_r^{\,\text{eff}}$ remains 
negative throughout the spacetime, and the violation of this single condition 
is sufficient to ensure the breakdown of the WEC. In Fig.~\ref{fig6}, we can 
observe that the radial DEC is violated for all values of $m$. On the other 
hand, the tangential DEC is violated only for $1/8 < m < 1/2$ at the throat, 
while it is satisfied for smaller values of $m$. These results clearly 
indicate that the violation of energy conditions is more pronounced in the 
radial direction compared to the tangential direction. Moreover, all the 
plotted quantities tend toward zero as $r$ increases, implying that the 
violations are confined to a finite region around the wormhole throat. Such 
localization of exotic matter is a desirable feature for physically viable 
wormhole models, as it minimizes the extent of energy condition violations 
in spacetime.

\begin{figure}[!h]
    \centerline{
    \includegraphics[scale=0.85]{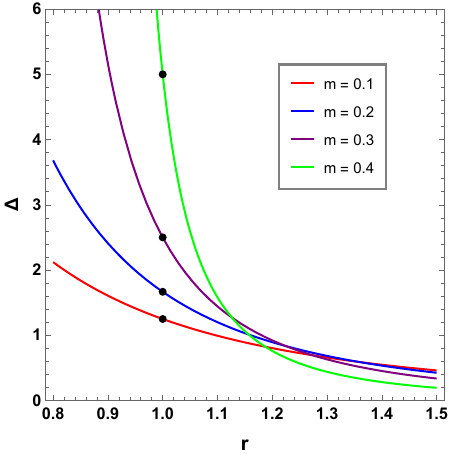}
    \hspace{1cm}
    \includegraphics[scale=0.875]{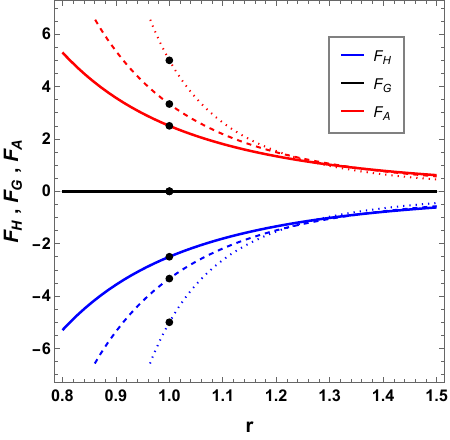}}
    \vspace{-0.2cm}
    \caption{Plot of $\Delta$ versus $r$ (left panel), and the profiles of 
$F_H$, $F_G$ and $F_A$ versus $r$ (right panel) with different values of the 
parameter $m$, throat radius $r_0 = 1$ and $\Phi(r)=0$. The thick, dashed and 
dotted curves on the right plot correspond to $m=0.1$, $m=0.2$ and $m=0.3$, respectively.}   
    \label{fig7}
\end{figure}

We now examine the anisotropy parameter $\Delta$ and the forces governing the 
equilibrium of the system, namely the hydrostatic force $F_H$, anisotropic 
force $F_A$, and gravitational force $F_G$. These quantities are given by
\begin{align}
\Delta & =
-\,\frac{(2m - 1)\, r_0 \left[6m(r - r_0) + r_0\right]}
{r^2 \left(4mr - 6mr_0 + r_0\right)^2},
\label{eq63}\\[8pt]
%
F_H & =
\frac{2(2m - 1)\, r_0 \left[6m(r - r_0) + r_0\right]}
{r^3 \left(4mr - 6mr_0 + r_0\right)^2},
\label{eq64}\\[8pt]
%
F_A & =
-\frac{2(2m - 1)\, r_0 \left[6m(r - r_0) + r_0\right]}
{r^3 \left(4mr - 6mr_0 + r_0\right)^2},
\label{eq65}\\[8pt]
%
F_G & = 0.
\label{eq66}
\end{align}
It is clear from Eqs.~\eqref{eq64} and \eqref{eq65} that $F_A = -\,F_H$ for 
all values of $m$. Consequently, the equilibrium condition 
$F_H + F_A + F_G = 0$ is satisfied identically without any contribution from 
the gravitational force.
Using Eqs.~\eqref{eq63} - \eqref{eq66}, we plot the anisotropy parameter 
$\Delta$ and the profiles of the hydrostatic force $F_H$, gravitational force 
$F_G$, and anisotropic force $F_A$ as functions of the radial coordinate $r$ 
for different values of the parameter $m$, with the throat radius fixed at 
$r_0 = 1$, as shown in Fig.~\ref{fig7}.

From the left panel of Fig.~\ref{fig7}, it is observed that the anisotropy 
parameter $\Delta$ remains positive throughout the spacetime for all values of 
$m$. This implies that the tangential pressure exceeds the radial pressure, 
i.e., $p_t^{\,\text{eff}} > p_r^{\,\text{eff}}$, resulting in a repulsive 
anisotropic force. The magnitude of $\Delta$ is maximum near the wormhole 
throat and decreases monotonically with increasing $r$, approaching zero at 
large radial distances. This indicates that the anisotropic effects are 
strongest near the throat and gradually diminish far from it. Furthermore, 
$\Delta$ increases with increasing values of the parameter $m$, showing that 
the parameter $m$ enhances the degree of anisotropy and strengthens the 
repulsive contribution required to sustain the wormhole structure.
The right panel of Fig.~\ref{fig7} shows the behavior of the hydrostatic, 
anisotropic, and gravitational forces. Since the redshift function is zero, 
the gravitational force vanishes identically, as indicated by the flat line 
along the horizontal axis. The hydrostatic force $F_H$ is negative throughout 
spacetime, implying an inward-directed force, whereas the anisotropic 
force $F_A$ is positive and acts outward. Thus, the wormhole configuration is 
maintained by the balance between the inward hydrostatic force and the outward 
anisotropic force. Moreover, the magnitudes of both $F_H$ and $F_A$ are larger 
near the throat and decrease with increasing $r$, indicating that the forces 
responsible for maintaining equilibrium are localized around the wormhole 
throat. This behavior becomes more pronounced for larger values of $m$, 
further confirming that the parameter $m$ controls the strength of the 
anisotropic support in the system.

\subsection{For $\boldsymbol{\Phi(r)=\log\left(1+\frac{r_0}{r}\right)}$ case}
\label{subsec0502}
%
%

\begin{figure}[!h]
    \centerline{
    \includegraphics[scale=0.8]{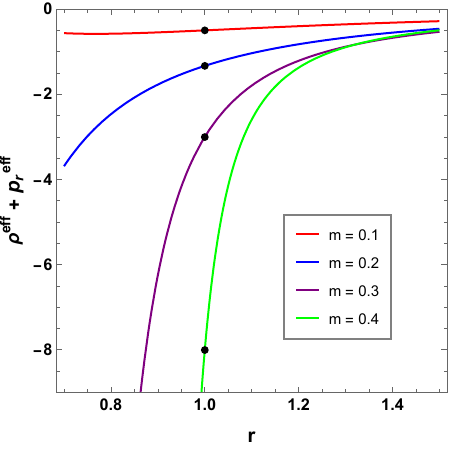}
    \hspace{1cm}
    \includegraphics[scale=0.81]{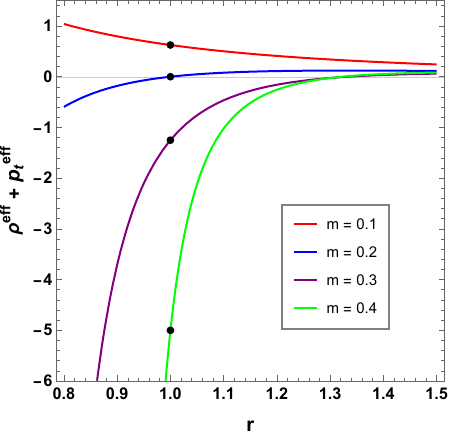}}
    \vspace{-0.2cm}
    \caption{Variation of $\rho^{\,\text{eff}} + p_r^{\,\text{eff}}$
    (left panel) and $\rho^{\,\text{eff}} + p_t^{\,\text{eff}}$ (right 
    panel)versus $r$ for different values of $m$ with redshift function $
    \Phi(r)=\log\left(1+r_0/r\right)$ and throat radius $r_0 = 1$.}
    \label{fig8}
\end{figure}

\begin{figure}[!h]
    \centerline{
    \includegraphics[scale=0.8]{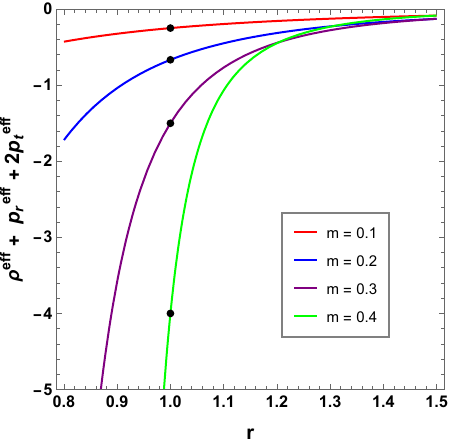}}
    \vspace{-0.2cm}
    \caption{Plot of $\rho^{\,\text{eff}} + p_r^{\,\text{eff}} + 2p_t^{\,
    \text{eff}}$ versus $r$ for different values of $m$ with redshift 
    function $\Phi(r)=\log\left(1+r_0/r\right)$ and throat radius $r_0 = 1$.}
    \label{fig9}
\end{figure}

\begin{figure}[!h]
    \centerline{
    \includegraphics[scale=0.8]{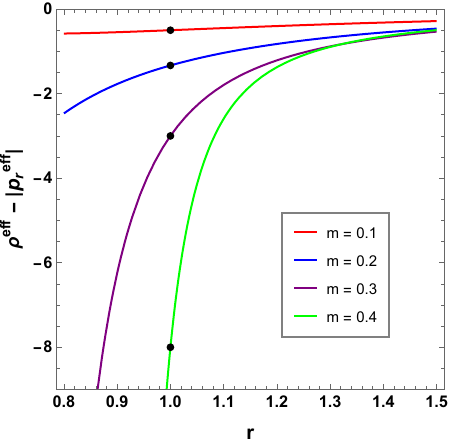}
    \hspace{1cm}
    \includegraphics[scale=0.81]{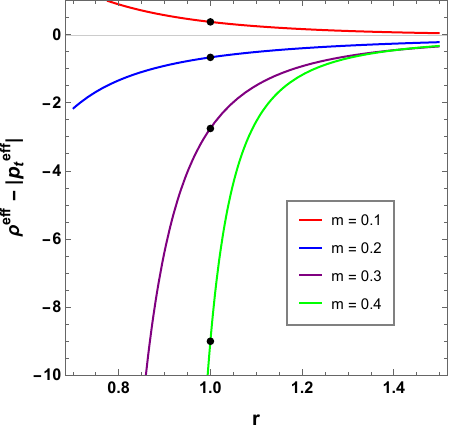}}
    \vspace{-0.2cm}
    \caption{Variation of $\rho^{\,\text{eff}} - \left|p_r^{\,\text{eff}}\right|
    $ (left panel) and $\rho^{\,\text{eff}} - \left|p_t^{\,\text{eff}}\right|$ 
	(right panel) with respect to $r$ for different values of $m$ with redshift 
	function $\Phi(r)=\log\left(1+r_0/r\right)$ and throat radius $r_0 = 1$.}
    \label{fig10}
\end{figure}

For this redshift function, from Eqs.~\eqref{eq51} and~\eqref{eq52}, we can 
obtain the specific forms of effective radial pressure and tangential pressure 
as
\begin{align}
p_r^{\,\text{eff}} & =
-\frac{r_0 \left(6 m r - 10 m r_0 + r + r_0\right)}
{r^2 (r + r_0)\left(4 m r - 6 m r_0 + r_0\right)}, \label{eq68}\\[8pt]
%
p_t^{\,\text{eff}} & = 
\frac{2 m r_0 \left[(6 m + 1) r^2 + (2 - 20 m) r r_0 + 2 (6 m - 1) r_0^2\right]}
{r^2 (r + r_0)\left(4 m r - 6 m r_0 + r_0\right)^2} . \label{eq69}
\end{align}
Using these two Eqs.~\eqref{eq68} and \eqref{eq69}, the specific expressions 
for the four energy conditions can be written as
\begin{equation}
p_r^{\,\text{eff}} =
-\frac{r_0 \left(6 m r - 10 m r_0 + r + r_0\right)}
{r^2 (r + r_0)\left(4 m r - 6 m r_0 + r_0\right)}, \label{eq68}
\end{equation}

\begin{equation}
p_t^{\,\text{eff}}  = 
\frac{2 m r_0 \left[(6 m + 1) r^2 + (2 - 20 m) r r_0 + 2 (6 m - 1) r_0^2\right]}
{r^2 (r + r_0)\left(4 m r - 6 m r_0 + r_0\right)^2} . \label{eq69}
\end{equation}
Using these two Eqs.~\eqref{eq68} and \eqref{eq69}, the specific expressions 
for the four energy conditions can be written as
\begin{align}
\rho^{\,\text{eff}} + p_r^{\,\text{eff}} & = 
-\frac{4 m r_0 \left[(6 m + 1) r^2 + (3 - 22 m) r r_0 + 2 (6 m - 1) r_0^2\right]}{r^2 (r + r_0)\left(4 m r - 6 m r_0 + r_0\right)^2}, \label{eq70}\\[8pt]
%
\rho^{\,\text{eff}} + p_t^{\,\text{eff}} & =
\frac{r_0 \left[2 m (6 m + 1) r^2 + \left(1 - 4 m (7 m + 1)\right) r r_0 + (1 - 6 m)^2 r_0^2 \right]}{r^2 (r + r_0)\left(4 m r - 6 m r_0 + r_0\right)^2},
\label{eq71}\\[8pt]
%
\rho^{\,\text{eff}} + p_r^{\,\text{eff}} & + 2  p_t^{\,\text{eff}} =
\frac{4 m (2 m - 1) r_0^2}
{r (r + r_0)\left(4 m r - 6 m r_0 + r_0\right)^2},
 \label{eq72}\\[8pt]
%
\rho^{\,\text{eff}} - \left|p_r^{\,\text{eff}}\right| & =
\frac{\left[4 m (3 m - 2) + 1\right] r_0^2}
{r^2 \left(4 m r - 6 m r_0 + r_0\right)^2}
- \left|
\frac{r_0 \left(6 m r + r - 10 m r_0 + r_0\right)}
{r^2 (r + r_0)\left(4 m r - 6 m r_0 + r_0\right)}
\right|, \label{eq73}\\[8pt]
%
\rho^{\,\text{eff}} - \left|p_t^{\,\text{eff}}\right| & =
\frac{\left[4 m (3 m - 2) + 1\right] r_0^2}
{r^2 \left(4 m r - 6 m r_0 + r_0\right)^2}
- 2 \left|
\frac{m r_0 \left[(6 m + 1) r^2 + (2 - 20 m) r_0 r + 2 (6 m - 1) r_0^2\right]}
{r^2 (r + r_0)\left(4 m r - 6 m r_0 + r_0\right)^2}
\right|. \label{eq74}
\end{align}
To examine the validity of the energy conditions, we plot expressions in 
Eqs.~\eqref{eq70} - \eqref{eq74} as functions of the radial coordinate $r$ in 
Figs.~\ref{fig8} - \ref{fig10}.

From the left panel of Fig.~\ref{fig8} it is seen that the radial NEC is 
violated throughout the entire interval $0 < m < 1/2$. In contrast, the 
tangential NEC, shown in the right panel of Fig.~\ref{fig8}, is satisfied only 
for smaller values of $m$ and becomes negative at the throat for $m > 1/5$. 
Furthermore, Fig.~\ref{fig9} indicates that SEC is violated near the throat 
for all values of $m$. DEC is also violated throughout the parameter range, as 
seen in Fig.~\ref{fig10}; in particular, $\rho^{\,\text{eff}} - |p_r^{\,\text{eff}}|$ remains negative near the throat for all $0 < m < 1/2$, while 
$\rho^{\,\text{eff}} - |p_t^{\,\text{eff}}|$ becomes negative for 
$1/7 < m < 1/2$. Therefore, the wormhole geometry is supported by exotic 
matter, with violations of all energy conditions in the vicinity of the 
throat. Moreover, all the plotted quantities decay to zero with increasing 
$r$, showing that the violations are confined to the vicinity of the wormhole 
throat, as expected for a physically acceptable wormhole geometry.

\begin{figure}[!h]
    \centerline{
    \includegraphics[scale=0.8]{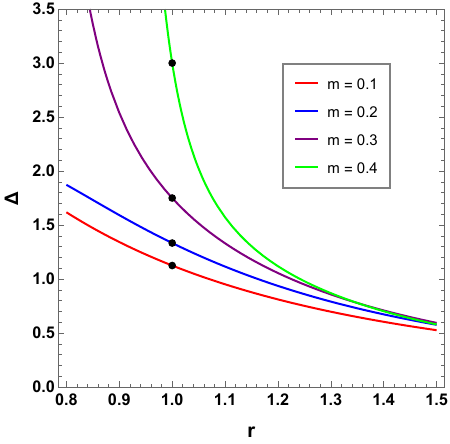}
    \hspace{1cm}
    \includegraphics[scale=0.8]{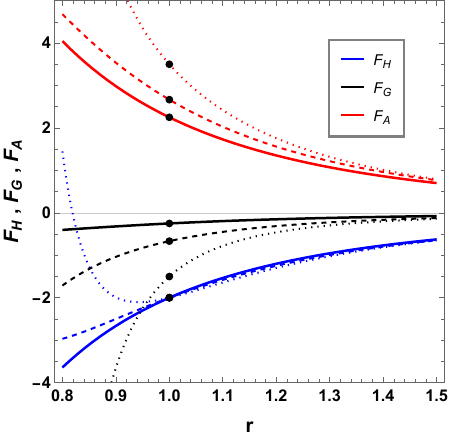}}
    \vspace{-0.2cm}
    \caption{Plot of $\Delta$ versus $r$ (left panel) and the radial profiles 
	of the hydrostatic force $F_H$, gravitational force $F_G$, and anisotropic 
	force $F_A$ (right panel) for different values of the parameter $m$ with 
	redshift function $\Phi(r)=\log\left(1+r_0/r\right)$ and throat
	radius $r_0 = 1$. The thick, dashed and dotted curves correspond to 
	$m=0.1$, $m=0.2$ and $m=0.3$, respectively.}
    \label{fig11}
\end{figure}

To examine the anisotropy parameter $\Delta$ and the forces governing the 
equilibrium of the system, namely the hydrostatic force $F_H$, anisotropic 
force $F_A$, and gravitational force $F_G$ for this redshift function, 
we write these quantities for this case as
\begin{align}
\Delta & =
\frac{r_0 \left[6 m (6 m + 1) r^2 + \left(4 (2 - 29 m) m + 1\right) r r_0 + \left(4 m (21 m - 5) + 1\right) r_0^2 \right]}
{r^2 (r + r_0)\left(4 m r - 6 m r_0 + r_0\right)^2},
\label{eq75}\\[8pt]
%
F_H & =
\frac{2 r_0 \left[
-2 (6 m^2 - 9 m + 1) r r_0^2
- 6 m (6 m + 1) r^3
+ \left(4 m (23 m - 3) - 1\right) r^2 r_0
+ \left(4 (4 - 15 m) m - 1\right) r_0^3
\right]}
{r^3 (r + r_0)^2 \left(4 m r - 6 m r_0 + r_0\right)^2},
\label{eq76}\\[8pt]
%
F_A & =
\frac{2 r_0 \left[6 m (6 m + 1) r^2 + \left(4 (2 - 29 m) m + 1\right) r_0 r + \left(4 m (21 m - 5) + 1\right) r_0^2 \right]}
{r^3 (r + r_0)\left(4 m r - 6 m r_0 + r_0\right)^2},
\label{eq77}\\[8pt]
%
F_G & = -\,\frac{4 m r_0^2 \left[(6 m + 1) r^2 + (3 - 22 m) r r_0 + 2 (6 m - 1) r_0^2 \right]}
{r^3 (r + r_0)^2 \left(4 m r - 6 m r_0 + r_0\right)^2}.
\label{eq78}
\end{align}
Eqs.~\eqref{eq76} - \eqref{eq78} shows that the equilibrium condition 
$F_H + F_A + F_G = 0$ is satisfied identically. Using Eqs.~\eqref{eq75} - 
\eqref{eq78}, we plot the anisotropy parameter $\Delta$ and the profiles of 
the hydrostatic force $F_H$, gravitational force $F_G$, and anisotropic force 
$F_A$ as functions of the radial coordinate $r$ for different values of the 
parameter $m$, with the throat radius fixed at $r_0 = 1$, as shown in 
Fig.~\ref{fig11}.

\textbf
It is observed from the left panel of Fig.~\ref{fig11} that the anisotropy 
parameter $\Delta$ remains positive throughout the spacetime for all 
considered values of $m$, indicating $p_t^{\,\text{eff}} > p_r^{\,\text{eff}}$.
This results in the presence of a repulsive anisotropic force, which plays a 
crucial role in supporting the wormhole structure against gravitational 
collapse. Moreover, the anisotropy remains finite at the throat, ensuring the 
regularity of the matter distribution. A notable feature of the profile is that 
$\Delta$ initially increases just outside the throat, attains a maximum value, 
and then gradually decreases with increasing $r$. The peak value of the 
anisotropy parameter increases with the parameter $m$, implying that higher 
values of $m$ enhance the strength of the anisotropic repulsion. 
For larger values of $r$, $\Delta$ decreases and approaches small positive values, indicating that the spacetime gradually tends toward a weakly anisotropic configuration.
This behavior suggests that the anisotropic effects are most significant near 
the wormhole throat and diminish asymptotically. The right panel of 
Fig.~\ref{fig11} illustrates that the anisotropic force $F_A$ remains positive 
throughout spacetime, representing a repulsive contribution, while the 
hydrostatic force $F_H$ and the gravitational force $F_G$ remain negative, 
indicating an attractive nature. The overall equilibrium condition 
$F_H + F_G + F_A = 0$ is satisfied at all radial distances. Near the wormhole 
throat, the anisotropic force $F_A$ dominates and balances the combined inward 
contributions of the hydrostatic and gravitational forces.
Further, all forces remain finite at the throat, ensuring the regularity of 
the configuration. As the radial coordinate increases, the magnitudes of all 
forces gradually decrease and tend toward zero, which indicates that the system 
approaches a weak-field regime at large distances. It is also observed that 
increasing the parameter $m$ enhances the magnitude of the anisotropic force 
and modifies the behavior of the hydrostatic force, thereby strengthening the 
internal force balance required to sustain the wormhole geometry.

\section{Scalar perturbation and quasinormal modes}\label{sec06}

Quasinormal modes (QNMs) are the characteristic damped oscillations of compact 
objects such as black holes and wormholes, which arises from perturbations of 
the spacetime geometry. These modes are described by complex frequencies, where 
the real part represents the oscillation frequency and the imaginary part 
determines the damping rate of the perturbative oscillations. The decay of 
these modes reflects the loss of energy from the system and eventually allowing 
it to settle down to a stable configuration. For comprehensive reviews on QNMs, 
one can see Refs.~\cite{a43,a44,a45,a46,a47,a48,a49,a50,a51,a52,a53,a54}.
To compute the QNMs, one typically introduces a test field propagating in 
the background spacetime, which acts as a probe of the underlying geometry. 
Depending on the nature of the perturbation, one may consider scalar, vector, 
or fermionic (Dirac) test fields~\cite{a55,a56,a57}. In this work, we consider 
scalar field perturbations. In our analysis, we neglect possible contributions 
from quantum corrections near the throat or from distant matter 
fields~\cite{a37,a58}. A key aspect in the analysis of wormhole perturbations 
is the behavior of the effective potential, which typically exhibits a peak 
near the throat. In standard radial coordinates, this feature is not always 
convenient to analyze. Therefore, it is useful to introduce the tortoise 
coordinate, which transforms the perturbation equation into a 
Schr\"{o}dinger-like form and allows for a clearer interpretation of the 
potential barrier.

Now, we consider a massless scalar field $\zeta$ around the wormhole 
spacetime. Considering that the scalar field has a negligible effect on 
spacetime, one can describe the QNMs of wormholes by the Klein-Gordon equation 
for curved spacetime as~\cite{a59}
\begin{equation}
\Box \zeta = \frac{1}{\sqrt{-g}} \partial_\mu \left( \sqrt{-g}\, g^{\mu\nu} \partial_\nu \zeta \right) = 0, \label{eq79}
\end{equation}
where $g=\det(g_{\mu\nu})$ is the determinant of the metric tensor, 
$g^{\mu\nu}$ is the contravariant form of the metric tensor, and 
$\partial_{\nu}$ is the partial derivative with respect to the coordinate 
systems. Using spherical harmonics, the scalar field can be decomposed as
\begin{equation}
\zeta(t, r, \theta, \phi) = e^{-i\omega t} \frac{\psi(r)}{r} Y^\ell_m(\theta, \phi), \label{eq80}
\end{equation}
where $\psi(r)$ is the radial part of the field, $ Y^\ell_m(\theta, \phi)$ are 
the spherical harmonics and $\omega$ represents the frequency of oscillation 
of the temporal part of the wave containing QNMs. Substituting Eq.~\eqref{eq80}
into Eq.~\eqref{eq79}, we obtain the Schr\"odinger-like wave equation as
\begin{equation}
\frac{\partial^2 \psi(r_*)}{\partial r_*^2} + \big[\omega^2 - V_s(r_*)\big] \psi(r_*) = 0, \label{eq81}
\end{equation}
where $r_*$ represents the tortoise coordinate and is defined by
\begin{equation}
\frac{dr_*}{dr} =  e^{-\Phi(r)} \sqrt{\frac{r}{r - b(r)}}, \label{eq82}
\end{equation}
and $V_s(r)$ is the effective potential, which is given by
\begin{equation}
V_s(r) = e^{2\Phi(r)}\! \left[ \frac{l(l+1)}{r^2} + \frac{b(r) - r b'(r)}{2r^3} + \frac{\Phi'(r)}{r}\left(1 - \frac{b(r)}{r}\right) \right]. \label{eq83}
\end{equation}
Here $l$ is referred to as the multipole number of the QNMs of 
the wormhole. The effective potential has its maximum value located at 
the throat of the wormhole $r = r_0$ and monotonically decay at both 
infinities, representing either two different universes or two distant regions 
within the same universe. Therefore, it is natural that the QNMs of wormholes 
satisfy the following boundary conditions \cite{a60}:
\begin{equation}
\Psi(r_*) \sim 
\begin{cases}
e^{-i \omega r_*}\!, & r_* \to +\infty, \\[5pt]
e^{+i \omega r_*}\!, & r_* \to -\infty.
\end{cases} \label{eq84}
\end{equation}
These conditions imply purely outgoing waves at both infinities.

\subsection{WKB Method with Pad\'e Approximation}\label{subsec0601}

The semi-analytical WKB approximation method was first applied to black hole 
perturbations by Schutz and Will in 1985~\cite{a61}. The formalism was 
subsequently extended to higher orders, with the third-order WKB method 
developed in Ref.~\cite{a62} and later improved to sixth and higher orders by 
Konoplya~\cite{a63}. It has also been shown that the accuracy of the WKB 
method can be further enhanced by employing Pad\'e approximants~\cite{a64}.

Here, we employ the sixth-order WKB method with Pad\'e approximation to 
compute the quasinormal frequencies associated with the scalar field 
perturbations for the redshift functions $\Phi(r) = 0$ and $\Phi(r) = 
\log\left(1+r_0/r\right)$ separately. The numerical uncertainty associated 
with the WKB approximation is estimated using a commonly adopted prescription 
in the literature~\cite{a63,a65,a66}, which is given by
\begin{equation}
\delta_6 = \frac{\left| \text{WKB}_7 - \text{WKB}_5 \right|}{2}, \label{eq85}
\end{equation}
where $\text{WKB}_7$ and $\text{WKB}_5$ denote the QNM frequencies obtained 
from the seventh and fifth-order WKB approximations, respectively. This 
quantity provides an estimate of the truncation error arising from the 
finite-order WKB expansion, with smaller values of $\delta_6$ indicating 
better convergence of the method.

\subsubsection{QNMs for $\mathit{\Phi(r) = 0}$ case}\label{subsec0602}

\begin{figure}[!h]
    \centerline{
    \includegraphics[scale=0.85]{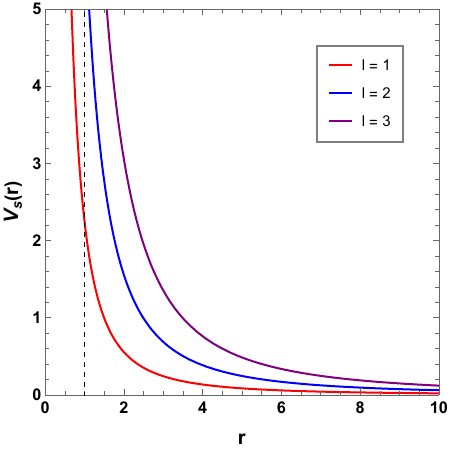}
    \hspace{1cm}
    \includegraphics[scale=0.85]{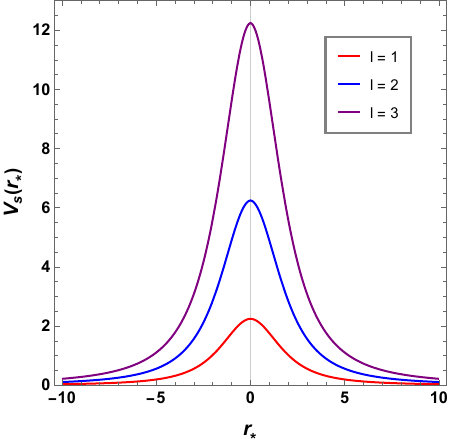}}
    \caption{Variation of effective potential $V_s(r)$ with respect to $r$ 
	(left panel) and $V_s(r_*)$ with respect to $r_*$ (right panel) for 
	different multipole numbers $l=1$, $2$, $3$ with $\Phi(r) = 0$, $m=0.1$ and 	
	$r_0=1$. The vertical dotted line in the left plot represents the position 
	of the throat of the wormhole.}
    \label{fig12}
\end{figure}

For the tideless condition, i.e., for the redshift function $\Phi(r)=0$ and 
the shape function given in Eq.~\eqref{eq49}, the effective potential in 
Eq.~\eqref{eq83} simplifies to
\begin{equation}
V_s(r)=\frac{l(l+1)}{r^2} + \frac{2(1 - 2m)m r r_0}{r^2\left(4mr - 6mr_0 + r_0\right)^2}.
\label{eq86}
\end{equation}
We plot the effective potential in terms of both the radial coordinate $r$ and 
the tortoise coordinate $r_*$ for different multipole number values $l=1$, 
$2$, $3$ with the model parameters fixed at $m=0.1$ and throat radius 
$r_0=1$, as shown in Fig.~\ref{fig12}.
The effective potential $V_s(r)$ remains positive over the entire domain and 
decreases smoothly as $r$ increases, eventually approaching zero at large 
distances. This suggests that far from the wormhole throat, spacetime 
behaves like a free-wave region. Near the throat ($r = r_0$), the potential 
rises sharply and in the range $r<r_0$, the potential tends to be singular at 
$r\rightarrow 0$. As this behavior of potential below the throat radius is 
unphysical, the tortoise coordinate transformation has been introduced to 
set the zero of the coordinate at the throat of the wormhole. Thus, the 
physically meaningful picture is therefore provided by the potential expressed 
in terms of the tortoise coordinate, $V_s(r_*)$, which remains regular across
all values of $\pm\,r_*$. In terms of $r_*$, the effective potential forms a 
single peak located at the wormhole throat and vanishes as 
$r_* \to \pm\, \infty$, consistent with the asymptotic structure of 
spacetime. The height of this peak increases with the multipole number $l$, 
reflecting the stronger centrifugal contribution for higher modes. At the 
same time, the potential becomes narrower, indicating that the perturbations 
are more tightly localized around the peak. This barrier-like profile is 
well-suited for the application of the WKB approximation and supports the 
existence of QNMs.

We now compute the fundamental QNM frequencies for different values of the 
multipole number using the Pad\'e-averaged WKB method, and the results are 
listed in Table~\ref{tab1}. It can be seen that the real part of the frequency 
increases steadily with the multipole number $l$, which means that higher 
multipole modes oscillate at higher frequencies. This trend is consistent with 
the increase in the height of the effective potential barrier for larger 
values of $l$. In contrast, the imaginary part of the frequency shows only a 
weak dependence on $l$ and remains nearly constant with a slight increase in 
magnitude. This indicates that the damping rate of the perturbations changes 
very little as $l$ increases. Since the imaginary part is negative for all 
modes, the perturbations decay with time. This indicates the stability of the 
wormhole under scalar perturbations. We also observe that the estimated error 
$\delta_6$ decreases significantly for higher multipole numbers, suggesting 
improved convergence of the WKB approximation. This is expected, as the WKB 
method generally performs better for potentials with sharper and more 
well-defined peaks, as seen in the effective potential profiles.
\begin{table}[!h]
\caption{Fundamental QNMs of scalar perturbation of the wormhole with zero 
redshift function for different multipole numbers with model parameter 
$m=0.1$ and throat radius $r_0=1$.}
\vspace{3pt}
\begin{tabular}{c c c c}
\hline\hline
Multipole number & Pad\'e averaged 6th-order WKB & $\delta_6$ \\
\hline
$l=1$ & $1.46926 - 0.247195i$ & $2.79428\times10^{-5}$ \\
$l=2$ & $2.48135 - 0.248985i$ & $2.51134\times10^{-6}$ \\
$l=3$ & $3.48664 - 0.249482i$ & $4.83498\times10^{-7}$ \\
$l=4$ & $4.48960 - 0.249687i$ & $1.39534\times10^{-7}$ \\
\hline\hline
\end{tabular}
\label{tab1}
\end{table}

\begin{figure}[!h]
    \centerline{
    \includegraphics[scale=0.85]{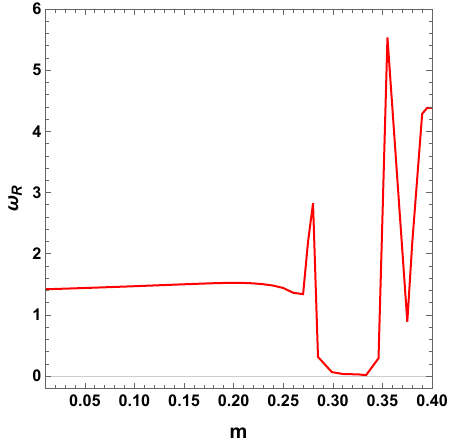}
    \hspace{1cm}
    \includegraphics[scale=0.9]{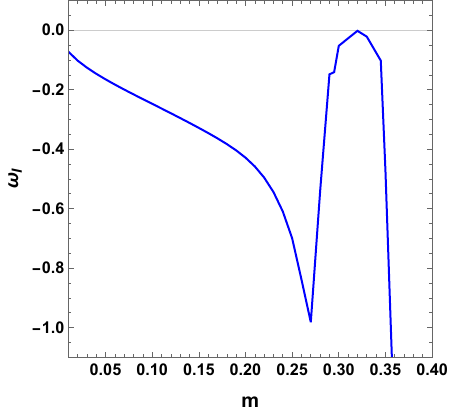}}
    \vspace{-0.2cm}
    \caption{Variation of real (left panel) and imaginary (right panel) parts 
of QNMs with respect to the model parameter $m$ within its allowed range 
for multipole number $l=1$ with redshift function $\Phi(r) = 0$ and throat
radius $r_0=1$.}
    \label{fig13}
\end{figure}

Fig.~\ref{fig13} illustrates how the fundamental quasinormal frequencies 
depend on the parameter $m$ for $l=1$ with $r_0=1$. Although the model allows 
the range $0<m<0.5$, we limit our analysis for QNMs to $0<m<0.25$. This is 
because, beyond $m \approx 0.25$, the QNMs begin to show abrupt 
oscillation behaviour, suggesting that the results are no longer reliable. 
This behavior may be linked to the decrease of the shape function as $m$ 
increases and significantly modifies the effective potential, and hence affects 
the validity of the WKB approximation.

\begin{figure}[!h]
    \centerline{
    \includegraphics[scale=0.85]{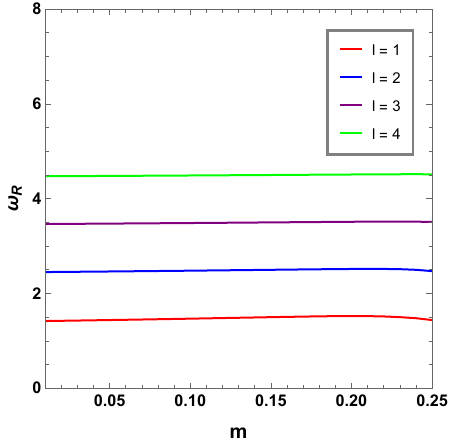}
    \hspace{1cm}
    \includegraphics[scale=0.9]{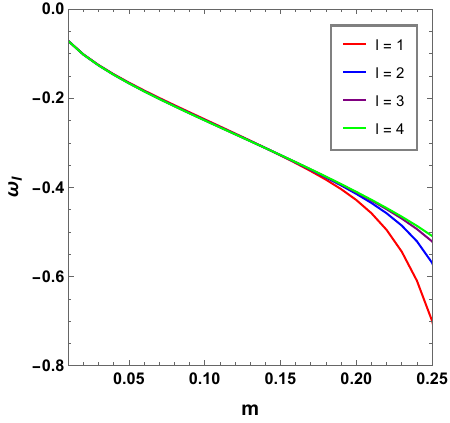}}
    \vspace{-0.2cm}
    \caption{Variation of real (left panel) and imaginary (right panel) parts 
of QNMs with respect to the model parameter $m$ within its range of values from 
$0$ to $0.25$ for multipole numbers $l=1$, $2$, $3$, $4$ with redshift 
function $\Phi(r) = 0$ and throat radius $r_0=1$.}
\label{fig14}
\end{figure}

Within the range $0<m<0.25$, both $\omega_R$ and $\omega_I$ vary smoothly with 
$m$, indicating a well-behaved quasinormal spectrum (see Fig.~\ref{fig14}). 
The imaginary part remains negative throughout, implying that the 
perturbations decay with time and indicating the stability of the wormhole 
under scalar perturbations. For these reasons, we restrict our QNM analysis 
to the parameter range where the effective potential is well behaved and the 
WKB method can be applied reliably.

\subsubsection{QMNS for $\mathit{\Phi(r) = \log\left(1+\frac{r_0}{r}\right)}$}
\label{subsec0603}

\begin{figure}[!h]
    \centerline{
    \includegraphics[scale=0.64]{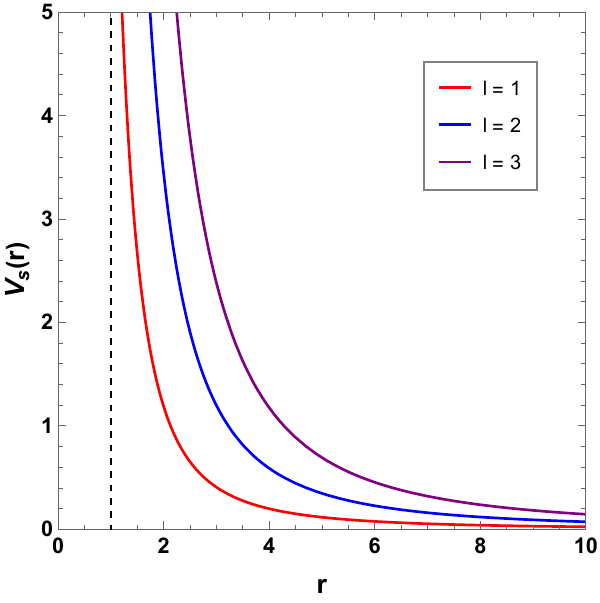}
    \hspace{1cm}
    \includegraphics[scale=0.64]{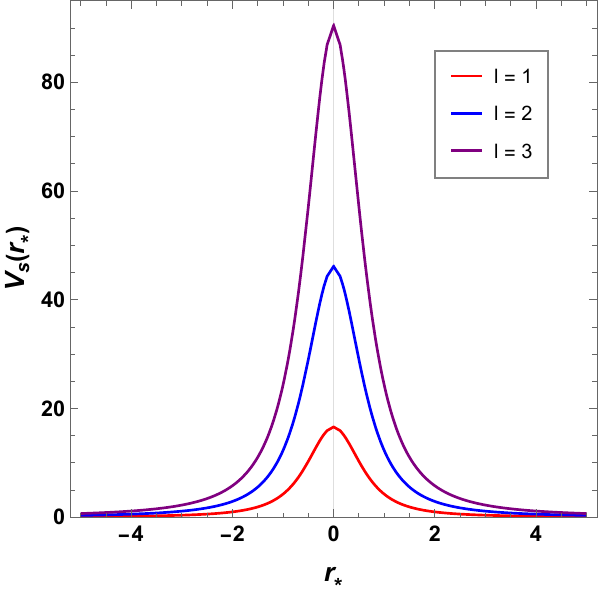}}
    \vspace{-0.2cm}
    \caption{Plots of effective potential $V_s(r)$ versus $r$ (left panel) and 
	$V_s(r_*)$ versus $r_*$ (right panel) for different multipole numbers 
	$l=1$, $2$, $3$ with $\Phi(r) = \log\left(1+r_0/r\right)$, $m=0.1$ and
	$r_0=1$. The vertical dotted line in the left plot represents the position 
	of the throat of the wormhole.}
    \label{fig15}
\end{figure}

For this redshift function 
and the shape function given in Eq.~(\ref{eq49}), the effective potential in 
Eq.~(\ref{eq83}) simplifies to
\begin{equation}
V_s(r)=
\left(1 + \frac{r_0}{r}\right)^2 \bigg[ \frac{l(l+1)}{r^2}
+
\frac{
r_0 \left(2 m (6 m + 1) r^2 + (1 - 4 m (7 m + 1)) r r_0 + (1 - 6 m)^2 r_0^2 \right)
}{
r^2 (r + r_0)\left(4 m r - 6 m r_0 + r_0\right)^2
} \bigg]. \label{eq87}
\end{equation}
We plot the effective potential in terms of both the radial coordinate $r$ and 
the tortoise coordinate $r_*$ for different multipole number values $l=1$, 
$2$, $3$, with the model parameters fixed at $m=0.1$ and throat radius 
$r_0=1$, as shown in Fig.~\ref{fig15}. As in the case of $\Phi(r) =0$, the 
potential expressed in terms of the radial coordinate, $V_s(r)$, remains 
positive throughout the domain, with sharply rising towards infinity as 
$r\rightarrow 0$ and decreasing monotonically with increasing 
$r$, approaching zero asymptotically. 
In the tortoise coordinate, the effective potential exhibits a single peak 
located near the wormhole throat and vanishes in the asymptotic regions 
$r_* \to \pm\, \infty$, consistent with the asymptotic structure of the 
spacetime. The height of the potential increases with the multipole number 
$l$, reflecting the contribution of the angular momentum term, while the 
barrier becomes narrower for higher $l$, indicating stronger localization of 
perturbations. The resulting single-peak structure supports a scattering 
interpretation and justifies the use of the WKB approximation for computing 
QNMs.

We compute the fundamental QNM frequencies for different values of the 
multipole number using the Pad\'e averaged WKB method. The corresponding 
results are presented in Table~\ref{tab2}. It is observed that the real part 
of the frequency increases monotonically with the multipole number $l$, 
indicating that higher multipole modes oscillate with higher frequencies. This 
behavior is consistent with the increase in the height of the effective 
potential barrier for larger values of $l$. The imaginary part of the 
frequency approaches an approximately constant value as $l$ increases, showing 
that the damping rate is only weakly dependent on the multipole number. Since 
the imaginary part remains negative for all values of $l$, the perturbations 
decay in time, indicating the stability of the wormhole under scalar 
perturbations. Furthermore, the estimated error $\delta_6$ decreases 
significantly with increasing $l$, demonstrating improved convergence of the 
WKB approximation for higher multipole modes. This is consistent with the fact 
that the WKB method becomes more accurate for potentials with sharper and 
narrower peaks. All these behaviours of calculated QNMs are similar to the
case of $\Phi(r)=0$ but with a higher magnitude of both amplitudes and
damping of the modes. 
\begin{table}[!h]
\caption{Fundamental QNMs of the wormhole with redshift function 
$\Phi(r) = \log\left(1+r_0/r\right)$ for scalar perturbation with model 
parameter $m=0.1$ and throat radius $r_0=1$.}
\vspace{3pt}
\begin{tabular}{c c c c}
\hline\hline
Multipole number & Pad\'e averaged 6th-order WKB & $\delta_6$ \\
\hline
$l=1$ & $2.90084 - 0.610280i$ & $2.79845\times10^{-4}$ \\
$l=2$ & $4.94057 - 0.611654i$ & $1.72659\times10^{-5}$ \\
$l=3$ & $6.95757 - 0.612005i$ & $2.91218\times10^{-6}$ \\
$l=4$ & $8.96700 - 0.612149i$ & $7.91038\times10^{-7}$ \\
\hline\hline
\end{tabular}
\label{tab2}
\end{table}

\begin{figure}[!h]
    \centerline{
    \includegraphics[scale=0.65]{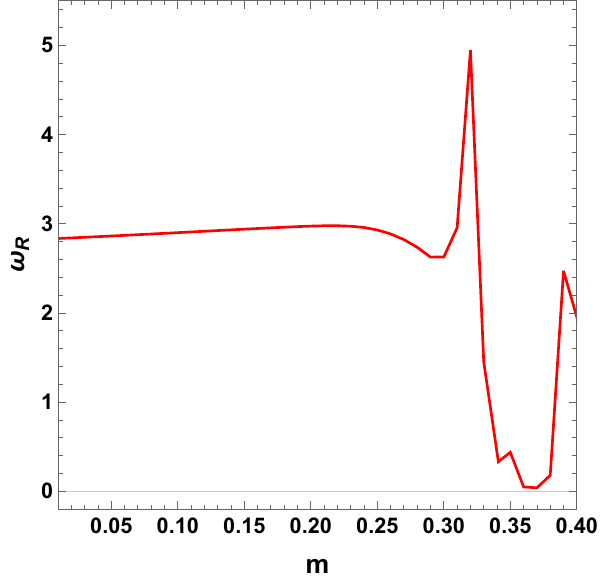}
    \hspace{1cm}
    \includegraphics[scale=0.69]{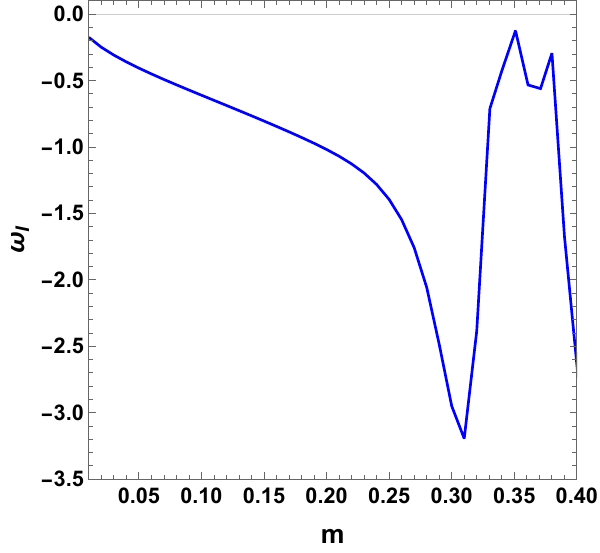}}
    \caption{Variation of real (left panel) and imaginary (right panel) parts 
of QNMs with respect to model parameter $m$ for multipole number $l=1$ 
with redshift function $\Phi(r) = \log\left(1+r_0/r\right)$ and throat radius 
$r_0=1$.}
    \label{fig16}
\end{figure}

Fig.~\ref{fig16} shows the dependence of the fundamental quasinormal 
frequencies on the parameter $m$ for $l=1$ with $r_0=1$. It is clear that 
although $0<m<0.5$ is the allowed range of the model parameter $m$, we must 
restrict the analysis to $0<m<0.30$ because abrupt oscillations appear in the 
frequencies for 
$m\gtrsim 0.30$, indicating a breakdown of the WKB approximation method and 
the loss of a well-defined single-peak effective potential. This behavior may 
be associated with the decrease of the shape function for increasing $m$, 
which significantly alters the structure of the effective potential as 
discussed in the case of $\Phi(r) = 0$. Thus, this restriction is consistent 
with the case for the zero redshift function.

\begin{figure}[!h]
    \centerline{
    \includegraphics[scale=0.65]{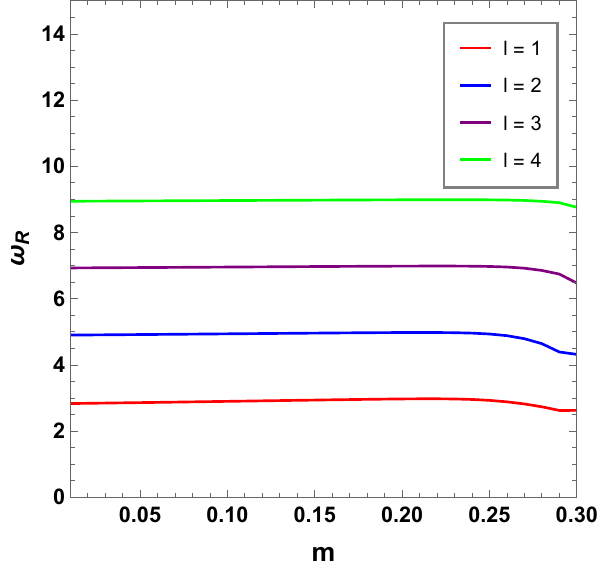}
    \hspace{1cm}
    \includegraphics[scale=0.67]{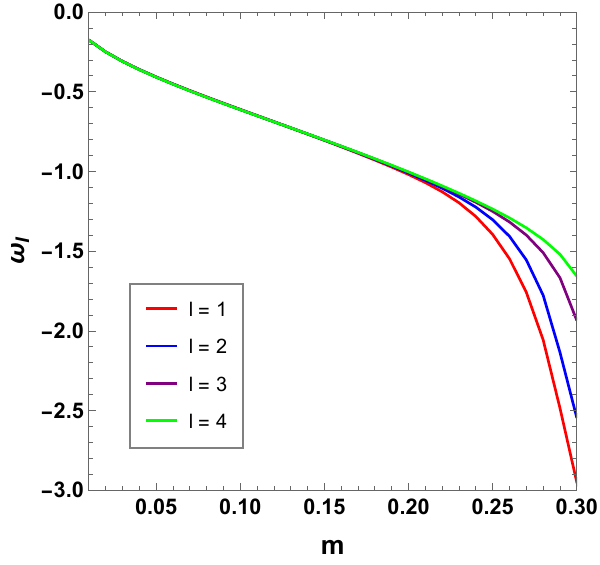}}
    \caption{Variation of real (left panel) and imaginary (right panel) parts 
of QNMs with respect to the model parameter $m$ within the range of values from 
$0$ to $0.25$ for multipole numbers $l=1$, $2$, $3$, $4$ with redshift 
function $\Phi(r) = \log\left(1+r_0/r\right)$ and throat radius $r_0=1$.}
    \label{fig17}
\end{figure}

Within the range $0<m<0.30$, both $\omega_R$ and $\omega_I$ vary smoothly with 
$m$, indicating a regular quasinormal spectrum (see Fig.~\ref{fig17}). The 
imaginary part remains negative throughout this region, implying that the 
perturbations decay in time and indicating the stability of the wormhole 
under scalar perturbations. Therefore, the quasinormal mode analysis is 
confined to the parameter range where the effective potential remains 
well-behaved, and the WKB method is applicable.

\subsection{Time-Domain Analysis}\label{sec07}

In this part of the section, we study the evolution of a scalar perturbation 
around the 
wormhole spacetime for both redshift functions. To study the evolution of 
the scalar field perturbation, we use the time-domain integration method 
described in Refs.~\cite{a67,a68}. Hence, introducing 
$\Psi(r_*,t) = \Psi(i \Delta r_*, j \Delta t) = \Psi_{i,j}$ and 
$V(r(r_*)) = V(r_*,t) = V_{i,j}$, Eq.~(\ref{eq79}) can be expressed as 
\begin{equation}
\frac{\Psi_{i+1,j} - 2\Psi_{i,j} + \Psi_{i-1,j}}{\Delta r_*^2}-\frac{\Psi_{i,j+1} - 2\Psi_{i,j} + \Psi_{i,j-1}}{\Delta t^2} - V_i \Psi_{i,j} = 0. \label{eq88}
\end{equation}
Considering
$\Psi(r_*, t) = \exp\left[-\dfrac{(r_* - k)^2}{2\sigma^2}\right]$ and 
$\left.\dfrac{\partial \Psi(r_*, t)}{\partial t}\right|_{t<0}\!\!\!\!\!\!= 0$, 
where $k$ and $\sigma$ are the median and width of the initial wave-packet, as 
initial conditions, the time evolution of the scalar field can be expressed as
\begin{equation}
\Psi_{i,j+1} = -\Psi_{i,j-1}
+ \left(\frac{\Delta t}{\Delta r_*}\right)^2
\left(\Psi_{i+1,j} + \Psi_{i-1,j}\right)
+ \left(2 - 2\left(\frac{\Delta t}{\Delta r_*}\right)^2 - V_i \Delta t^2 \right)\Psi_{i,j}. \label{eq89}
\end{equation}
We choose the condition $\Delta t/\Delta r_*< 1$ to satisfy the 
Von Neumann stability condition during the numerical calculations. The obtained
time domain profiles are fitted using the Levenberg–Marquardt algorithm
\cite{a69,a70,a71}. From the fitting, we extract the fundamental QNM 
frequencies of the wormhole spacetime with both redshift functions. To 
quantify the agreement between the WKB approximation and the time-domain 
results, we compute the absolute difference between the QNM frequencies as
\begin{equation}
\delta_{\text{QNM}} = \frac{\left| \text{QNM}_{\text{WKB}} - \text{QNM}_{\text{TDM}} \right|}{2}, \label{eq90}
\end{equation}
where subscript TDM stands for time-domain.
The coefficient of determination is computed as
\begin{equation}
R^2 = 1 - \frac{\sum \left( \Psi_{\text{TDM}} - \Psi_{\text{Fit}} \right)^2}{\sum \left( \Psi_{\text{TDM}} - \bar{\Psi}_{\text{TDM}} \right)^2}, \label{eq91}
\end{equation}
where $\Psi_{\text{TDM}}$ is the amplitude of the original time-domain 
waveform, $\Psi_{\text{Fit}}$ is the amplitude of the fitted waveform and 
$\bar{\Psi}_{\text{TDM}}$ is the mean value of the original time-domain 
waveform.

\subsubsection{For $\mathit{\Phi(r)=0}$ case}\label{subsec0701}

\begin{figure}[!h]
    \centering
    \includegraphics[scale=0.45]{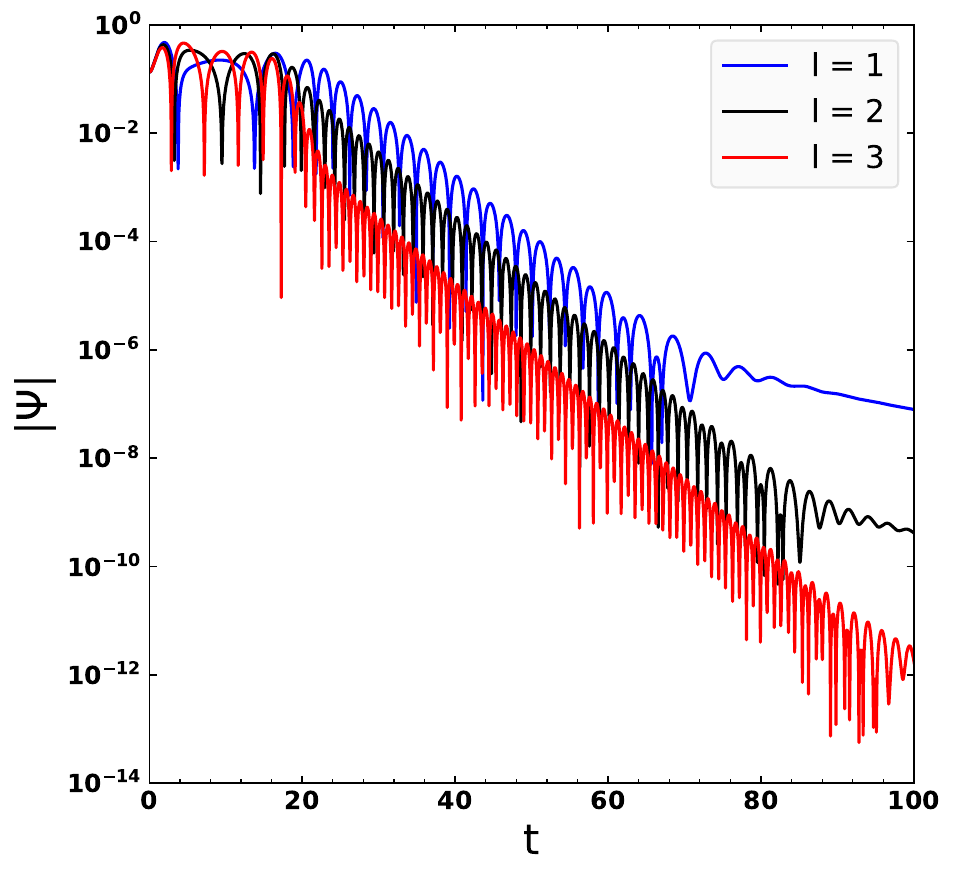}
    \vspace{-0.3cm}
    \caption{Time-domain evolution of the scalar perturbation $|\Psi(t)|$ 
in the wormhole spacetime for multipole numbers $l=1$, $2$, $3$, 
model parameter $m=0.1$, throat radius $r_0=1$ and redshift function 
$\Phi(r) = 0$.}
    \label{fig18}
\end{figure}

For this case of zero redshift function, i.e., for the tideless state of 
the wormhole, the time evolution of scalar perturbations for multipole numbers 
$l=1$, $2$, $3$ is shown in Fig.~\ref{fig18} for $m=0.1$ and $r_0=1$. The 
waveforms exhibit the characteristic behavior of perturbations, consisting of 
an initial transient phase followed by a quasinormal ringing stage and 
eventual decay. During the quasinormal ringing phase, the signal is dominated 
by damped oscillations. It is clearly observed that the oscillation frequency 
increases with the multipole number $l$, with higher $l$ modes oscillating 
more rapidly. This feature is in direct agreement with the real part of the 
quasinormal frequencies reported in Table~\ref{tab1}, where $\omega_R$ 
increases monotonically with $l$. In addition, the decay rates of the 
perturbations for different $l$ values are nearly identical, as indicated by 
the approximately parallel behavior of the curves in the plot. 
This is consistent with the imaginary part of the WKB results, which remains 
almost constant ($\omega_I \approx -\,0.27$) across different multipole 
numbers. Overall, the time-domain results are in good qualitative agreement 
with the WKB approximation, supporting the reliability of the computed QNMs 
for this wormhole configuration.
\begin{figure}[!h]
    \centerline{
    \includegraphics[scale=0.45]{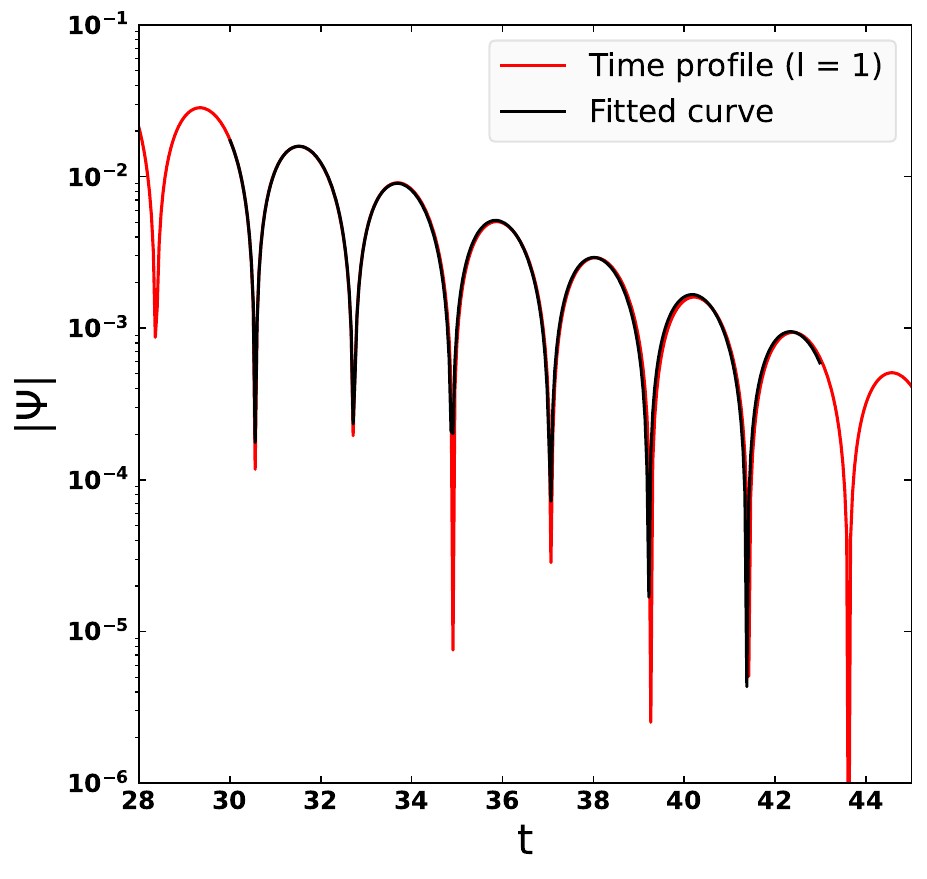}
    \hspace{0.5cm}
    \includegraphics[scale=0.45]{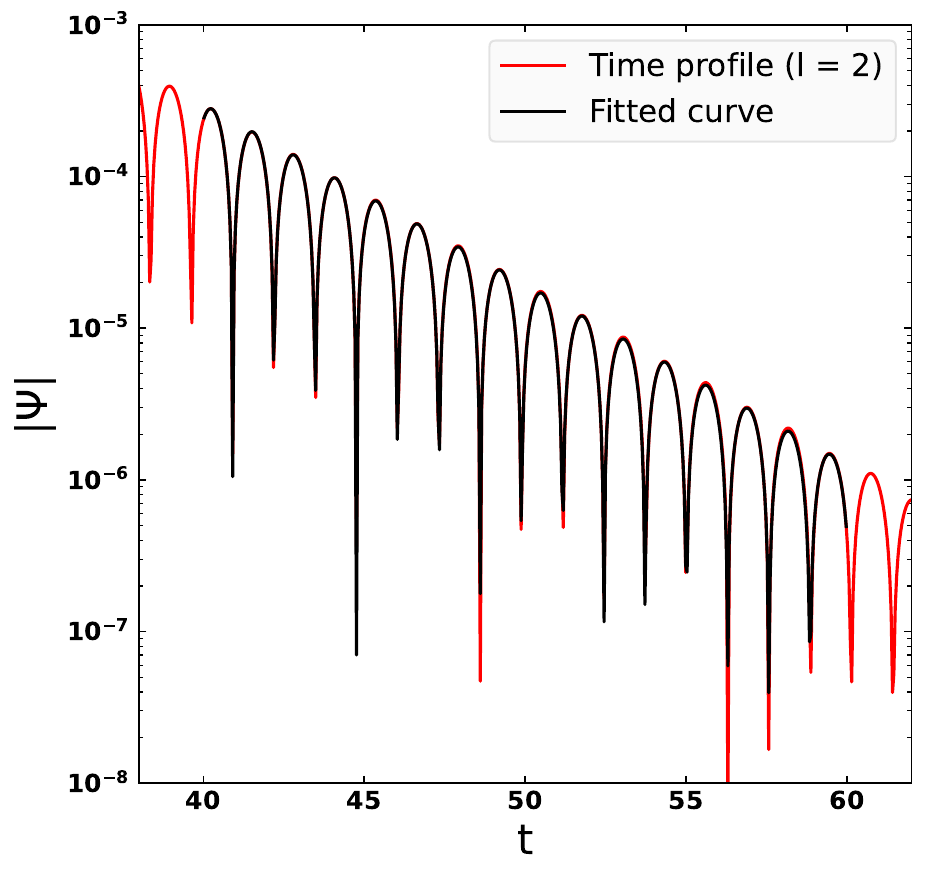}}
    \vspace{0.3cm}
    \centerline{
    \includegraphics[scale=0.45]{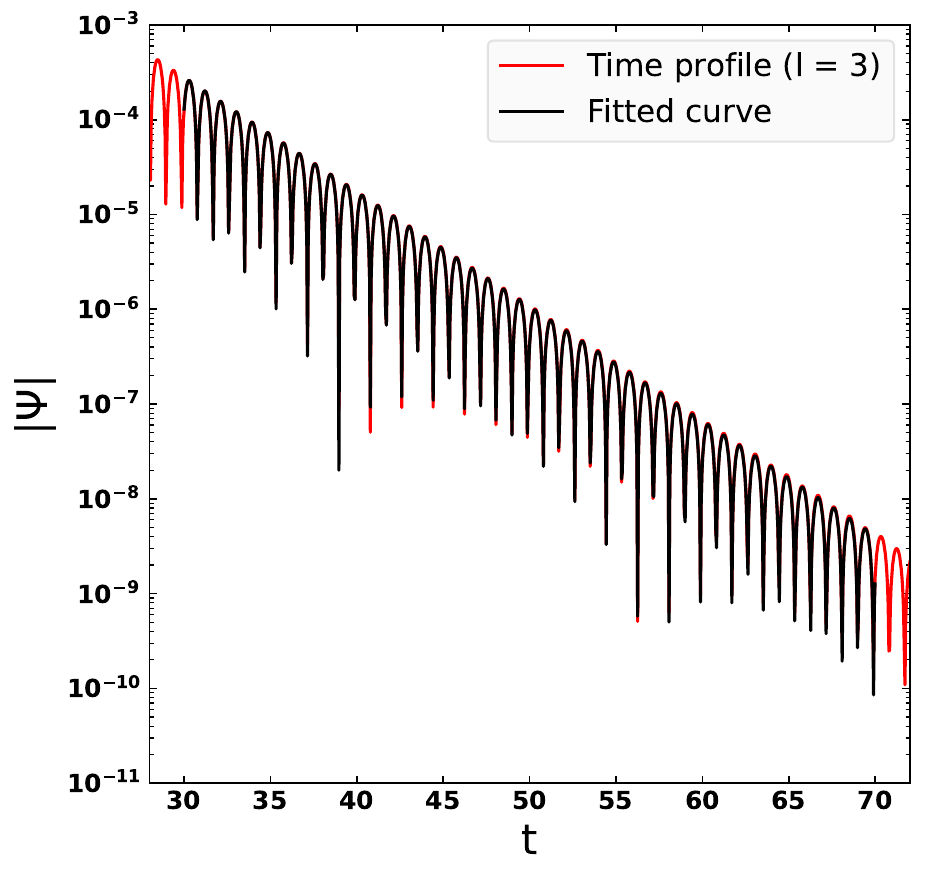}}
    \vspace{-0.3cm}
    \caption{Fitting of the time-domain profile of the wormhole to estimate 
its scalar QNMs for multipole numbers $l=1$, $2$, $3$, model parameter 
$m=0.1$, throat radius $r_0=1$ and redshift function $\Phi(r) = 0$.}
    \label{fig19}
\end{figure}

Fig.~\ref{fig19} shows the fitting of the time-domain profiles. 
Table~\ref{tab3} summarizes the fitted results in comparison with the 
corresponding WKB results. The agreement between the values of QNMs obtained 
from the WKB method and the fitting of time-domain curves, together with the 
high values of the coefficient of determination ($R^2 \sim 1$), indicates that 
the ringdown phase is well described by a single damped mode. However, we 
observe that the WKB method systematically overestimates the real part of the 
frequency and underestimates the damping rate compared to the time-domain 
results. Moreover, the deviation between the two methods increases with the 
multipole number $l$, indicating that the accuracy of the WKB approximation 
deteriorates for higher angular momentum modes in this spacetime.
\begin{table}[!h]
\caption{Fundamental QNMs of the wormhole with redshift function $\Phi(r) = 0$ 
for scalar perturbations, obtained from time-domain profiles in comparison 
with those from the WKB method with model parameter $m=0.1$ and throat radius 
$r_0=1$.}
\vspace{3pt}
\begin{tabular}{c c c c c}
\hline\hline
Multipole number & 6th-order WKB QNMs & Time-Domain QNMs & $R^2$ & $\delta_{\text{QNM}}$ \\
\hline
$l=1$ & $1.46926 - 0.247195i$ & $1.45226 - 0.260293i$ & $0.999759$ & $0.0107309$ \\
$l=2$ & $2.48135 - 0.248985i$ & $2.45188 - 0.272877i$ & $0.999976$ & $0.0189687$ \\
$l=3$ & $3.48664 - 0.249482i$ & $3.45318 - 0.278424i$ & $0.999990$ & $0.0221187$\\
\hline\hline
\end{tabular}
\label{tab3}
\end{table}

\subsubsection{For $\mathit{\Phi(r)=\log\left(1+\frac{r_0}{r}\right)}$ case}
\label{subsec0702}

\begin{figure}[!h]
    \centering
    \includegraphics[scale=0.45]{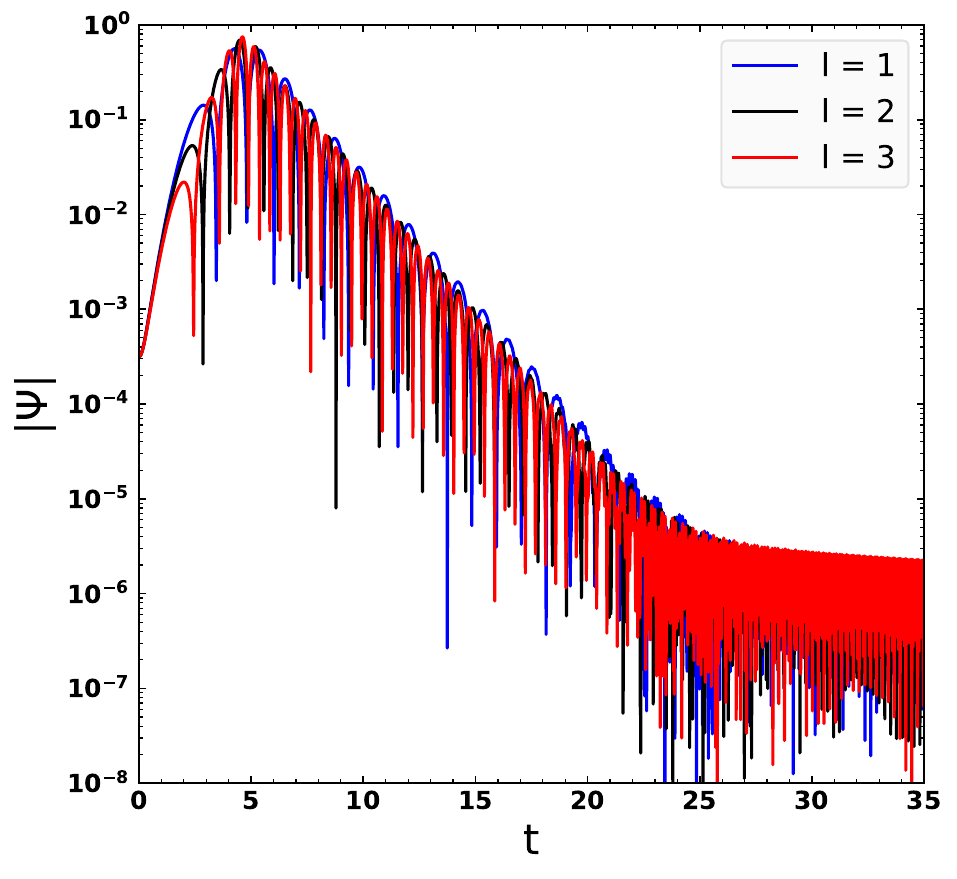}
    \vspace{-0.3cm}
    \caption{Time-domain evolution of the scalar perturbation $|\Psi(t)|$ of
the wormhole spacetime with the redshift function 
$\Phi(r)=\log\left(1+r_0/r\right)$ for multipole numbers $l=1$, $2$, $3$, 
model parameter $m=0.1$, and throat radius $r_0=1$.}
    \label{fig20}
\end{figure}

For this redshift function, 
the time-domain evolution of scalar perturbations for multipole numbers 
$l=1$, $2$, $3$ is shown in Fig.~\ref{fig20} for $m=0.1$ and $r_0=1$. The 
profiles clearly exhibit the standard three-stage behavior consisting of an 
initial transient phase, followed by a quasinormal ringing phase and late-time 
noises. With increasing multipole number $l$, the oscillation frequency 
increases while the damping rate slightly decreases, resulting in longer-lived 
oscillations. This behavior suggests that higher multipole modes are less 
dissipative and tend to dominate the signal at intermediate times. The overall 
decay of the perturbation amplitude indicates that the wormhole configuration 
remains stable under scalar perturbations for the chosen set of parameters. 
Furthermore, the presence of a logarithmic redshift function modifies the 
effective potential, leading to less damped and more sustained oscillations, 
which is reflected in the ringing phase compared to the zero redshift case.

\begin{figure}[!h]
    \centerline{
    \includegraphics[scale=0.45]{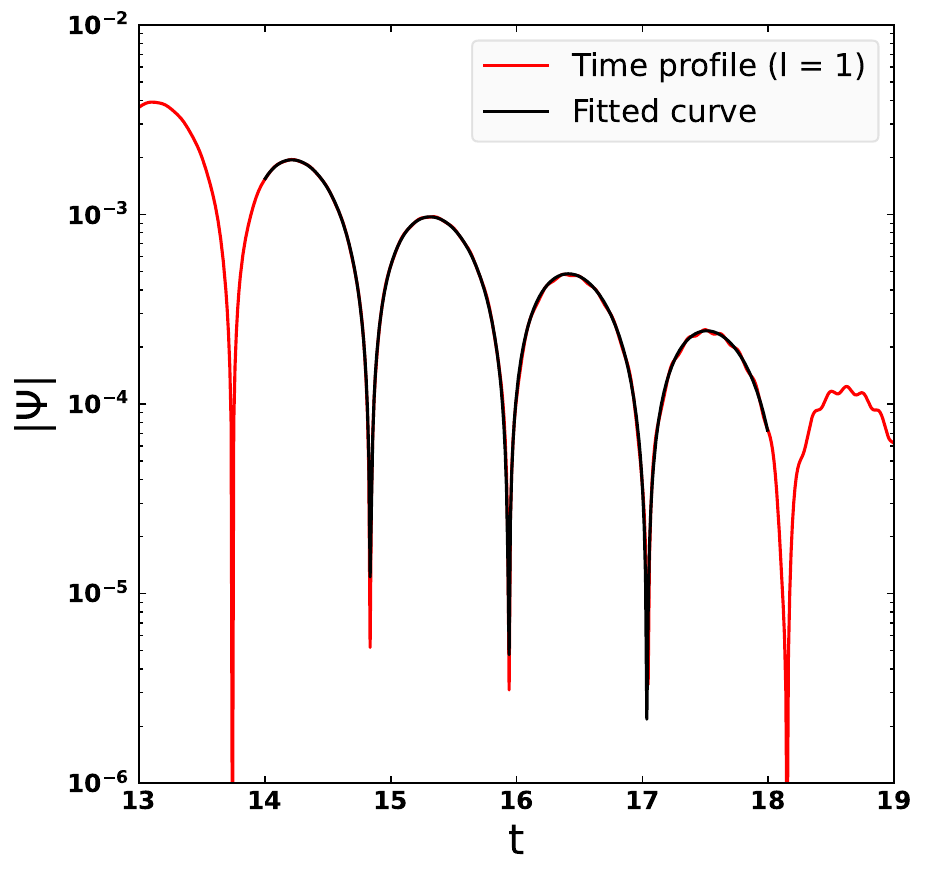}
    \hspace{0.5cm}
    \includegraphics[scale=0.45]{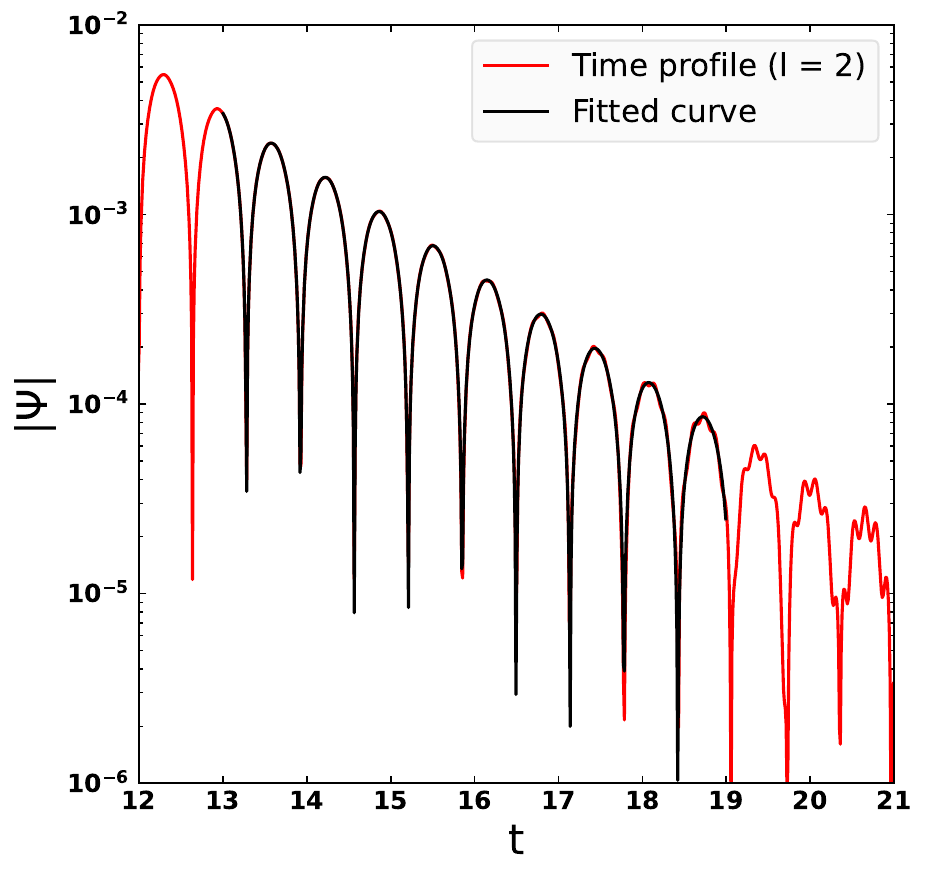}}
    \vspace{0.3cm}
    \centerline{
    \includegraphics[scale=0.45]{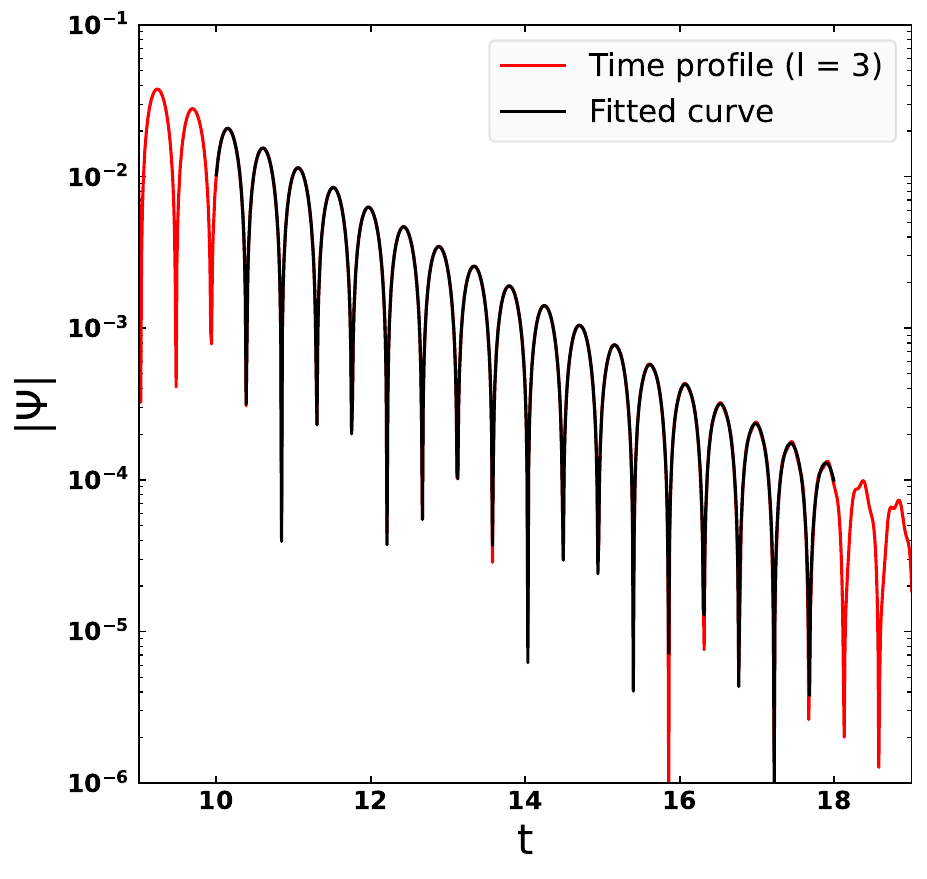}}
    \vspace{-0.3cm}
    \caption{Fitting of the time-domain profile of scalar perturbation of the
wormhole spacetime with logarithmic redshift function to estimate the QNMs 
for multipole numbers $l=1$, $2$, $3$, model parameter $m=0.1$, and throat 
radius $r_0=1$.}
    \label{fig21}
\end{figure}

Fig.~\ref{fig21} shows the fitting of the time-domain profiles of the scalar
perturbations of the wormhole with the logarithmic redshift function using 
the Levenberg-Marquardt algorithm for different multipole numbers. From the 
fitting, we extract the fundamental QNM frequencies of the wormhole spacetime 
with the considered redshift function and the results are summarized in 
Table~\ref{tab4} in comparison with those obtained from the WKB method for the
corresponding situation. The fitted curves exhibit excellent agreement with 
the numerical time-domain curves, accurately capturing the quasinormal ringing 
phase, as evident from the high values of the coefficient of determination 
$R^2 \sim 1$.
It is observed that the real part of the frequency increases with the 
multipole number $l$, indicating higher oscillation frequencies for higher 
modes, while the imaginary part remains negative, confirming the stability of 
the configuration under scalar perturbations. A comparison with the 6th-order 
WKB results shows good overall agreement. However, a systematic deviation is 
observed in the imaginary part, where the time-domain method predicts slightly 
larger damping rates. This discrepancy is quantified by the relative 
difference $\delta_{\text{QNM}}$, which remains below $\sim 0.04$ for all 
considered modes, indicating the reliability of both approaches. The deviation 
can be attributed to the limitations of the WKB approximation, especially in 
accurately capturing the damping behavior for higher $l$.
\begin{table}[!h]
\caption{Fundamental QNMs of the wormhole with redshift function $\Phi(r) = 
\log\left(1+r_0/r\right)$ for scalar perturbation with model parameter 
$m=0.1$ and throat radius $r_0=1$.}
\vspace{3pt}
\begin{tabular}{c c c c c}
\hline\hline
Multipole number & 6th-order WKB QNMs & Time-Domain QNMs & $R^2$ & $\delta_{\text{QNM}}$ \\
\hline
$l=1$ & $2.90084 - 0.610280i$ & $2.85928 - 0.629986i$ & $0.999962$ & $0.0230008$\\
$l=2$ & $4.94057 - 0.611654i$ & $4.88868 - 0.646533i$ & $0.999974$ & $0.0312592$\\
$l=3$ & $6.95757 - 0.612005i$ & $6.89794- 0.656168i$ & $0.999998$ & $0.0371024$\\
\hline\hline
\end{tabular}
\label{tab4}
\end{table}

\section{Conclusion}\label{sec08}

In this work, we have investigated traversable wormhole geometries within the 
framework of $f(Q)$ gravity by adopting a power-law form $f(Q)=\gamma(-Q)^m$. 
Starting from the modified field equations, we constructed a class of static 
and spherically symmetric wormhole solutions supported by an anisotropic fluid 
distribution and analyzed their physical viability through geometric, energy, 
and stability considerations. By imposing an EoS relating the 
radial pressure and energy density, we obtained an analytic form of the shape 
function, and identified the parameter range $0<m<1/2$ for which the 
flare-out and asymptotic flatness conditions are satisfied. This range 
corresponds to an effective EoS parameter $-1<\omega<-1/3$, indicating that 
the wormhole is supported by the effective geometry that mimics exotic but 
non-phantom matter in the quintessence regime. The embedding analysis in 
Section \ref{sec04} further confirmed that the resulting geometry represents a 
well-defined traversable wormhole with a smooth throat connecting two 
asymptotically flat regions.

The behavior of the effective energy density and pressures reveals that 
violations of the energy conditions are generally localized near the wormhole 
throat. In particular, the null and weak energy conditions are violated in 
the radial direction throughout spacetime, while tangential violations occur 
only within restricted parameter ranges. This localization of mimicking exotic 
matter is a desirable feature, as it minimizes unphysical behavior away from 
the throat. The anisotropy parameter remains positive in all cases, indicating 
that the tangential pressure exceeds the radial pressure and provides a 
repulsive contribution essential for maintaining the wormhole structure. The 
stability of the configuration was examined using the generalized TOV 
equation. For the zero redshift function or the tideless condition, 
equilibrium is achieved through an exact balance between hydrostatic and 
anisotropic forces, with no contribution from the gravitational term. In the 
case of a logarithmic redshift function, all three forces contribute, and the 
anisotropic force plays a dominant role near the throat by counteracting the 
combined inward pull of hydrostatic and gravitational forces. In both 
scenarios, the forces remain finite and decay asymptotically, ensuring a 
physically consistent equilibrium configuration.

To probe the dynamical stability of the wormhole, we analyzed scalar 
perturbations and computed the QNMs using the sixth-order WKB method with 
Pad\'e approximation as shown in Tables~\ref{tab1} and \ref{tab2}. The 
effective potential exhibits a single-peak barrier structure in the tortoise 
coordinate, justifying the applicability of the WKB approach. The resulting 
QNM frequencies show that the real part increases with the multipole number, 
while the imaginary part remains negative, confirming that perturbations decay 
with time. This indicates that the wormhole solutions are stable under scalar 
perturbations within the considered parameter range. However, we observe that 
the WKB method becomes unreliable for larger values of $m$, where the 
effective potential loses its well-defined peak structure due to violation of
the throat condition for such values of $m$. We further complemented the 
frequency-domain analysis with time-domain simulations of scalar 
perturbations. The evolution profiles exhibit the standard sequence of 
initial burst, quasinormal ringing, and late-time decay. The extracted 
frequencies from the time-domain fitting show good agreement with the WKB 
results as seen in Tables~\ref{tab3} and \ref{tab4}. However, small but 
systematic deviations appear in the damping rates. These discrepancies 
highlight the limitations of the WKB approximation method in accurately 
capturing the imaginary part of the frequencies, especially for higher 
multipole numbers.

Overall, our results demonstrate that $f(Q)$ gravity provides a viable 
framework for constructing traversable wormholes supported by anisotropic 
matter distribution with controlled violations of the energy conditions. The 
solutions are geometrically consistent, mechanically stable, and dynamically 
stable under scalar perturbations within a well-defined parameter range. The 
dependence of the wormhole properties on the model parameter $m$ further 
reveals that the alternative gravity sector plays a crucial role in regulating 
both the geometric structure and the physical viability of the solutions. 
Future work may extend this analysis to other forms of $f(Q)$, including 
different types of perturbations, such as electromagnetic or Dirac fields, and 
explore observational signatures that could distinguish such wormhole 
geometries from black hole spacetimes.

\section*{Acknowledgments}

UDG is thankful to the Inter-University Centre for Astronomy and Astrophysics 
(IUCAA), Pune, India, for the Visiting Associateship of the institute.

\appendix
\section{Calculations of Non-Metricity Tensor and Scalar}

For the wormhole metric~\eqref{eq20}, Eq.~\eqref{eq2} gives,
\begin{align}
Q_{rtt} & = -2\Phi'(r)e^{2\Phi(r)}, \nonumber \\[5pt]
Q_{rrr} & = \frac{r b'(r)-b(r)}{(r-b(r))^2}, \nonumber \\[5pt]
Q_{r\theta\theta} & = 2\,r, \nonumber \\[5pt]
Q_{r\phi\phi} & = 2\,r\sin^2\theta, \nonumber \\[5pt]
Q_{\theta\phi\phi} & = 2\,r^2\sin\theta\cos\theta.
\label{eqA1}
\end{align}
Hence, using equations in~(\ref{eqA1}), the non-metricity vector components 
from Eq.~\eqref{eq6} can be obtained as
\begin{align}
Q_{r} & = 2\Phi'(r) + \frac{r b'(r)-b(r)}{r(r-b(r))} + \frac{4}{r}, \nonumber \\
Q_{\theta} & = 2\cot\theta, \nonumber \\
\tilde{Q}_{r} & = \frac{r b'(r)-b(r)}{r(r-b(r))}, \nonumber \\
\tilde{Q}_{\theta} & = 0.
\label{eqA2}
\end{align}
Next, using Eqs.~\eqref{eqA2}, the components of superpotential tensor 
\eqref{eq8} can be found as
\begin{align}
P^{r}{}_{tt} & = -\frac{1}{r}\, e^{2\Phi(r)} \left(1-\frac{b(r)}{r}\right), \nonumber \\
P^{r}{}_{rr} & = 0, \nonumber \\
P^{r}{}_{\theta\theta} & = \left(1-\frac{b(r)}{r}\right)\left(\frac{r^2 \Phi'(r)}{2} + \frac{r}{2}\right),\nonumber \\
P^{r}{}_{\phi\phi} & = \left(1-\frac{b(r)}{r}\right)\left(\frac{r^2 \Phi'(r)}{2} + \frac{r}{2}\right)\sin^2\theta, \nonumber \\
P^{\theta}{}_{\phi\phi} & = 0.
\label{eqA3}
\end{align}
Therefore, finally the non-metricity scalar $Q$ for the wormhole~\eqref{eq7} 
takes the form:
\begin{align}
Q & = -\left(Q_{rtt}P^{rtt} + Q_{rrr}P^{rrr} + Q_{r\theta\theta}P^{r\theta\theta} + Q_{r\phi\phi}P^{r\phi\phi} + Q_{\theta\phi\phi}P^{\theta\phi\phi}\right), \nonumber \\[8pt]
& = -\frac{2}{r} \bigg[1-\frac{b(r)}{r}\bigg] \bigg[2\Phi'(r)+\frac{1}{r}\bigg].
\label{eqA4}
\end{align}

\end{document}